\documentclass[12pt,a4paper,english]{article}
\pdfoutput=1

\usepackage[bindingoffset=0.3cm,textheight=22.5cm,hdivide={2.7cm,*,2.7cm}, vdivide={*,22cm,*}]{geometry}
\usepackage[bookmarksnumbered=true,breaklinks=true]{hyperref}

\usepackage{amsmath,amsfonts,amssymb,babel,slashed,graphicx}

\usepackage[numbers,square,comma,sort&compress]{natbib}

\IfFileExists{dsfont.sty}
	{\usepackage{dsfont}
         \newcommand{\id}{\mathds{1}}}
	{\typeout{Package dsfont.sty was not found, using alternative macros.}
         \let\mathds=\mathbb
         \newcommand{\id}{\mbox{1 \kern-.59em {\rm l}}}}
\let\one=\id

\makeatletter

         \@addtoreset{equation}{section}
\makeatother

\newcommand{\nocontentsline}[3]{}
\newcommand{\tocless}[3]{\bgroup\let\addcontentsline=\nocontentsline#1{#2}#3\egroup}

\newcommand{\qed}{\nobreak \ifvmode \relax \else
      \ifdim\lastskip<1.5em \hskip-\lastskip
      \hskip1.5em plus0em minus0.5em \fi \nobreak
      \vrule height0.75em width0.5em depth0.25em\fi}

\newcommand{\be}{\begin{equation}}
\newcommand{\ee}{\end{equation}}
\newcommand{\eq}[1]{(\ref{#1})}

\def\nn{\nonumber}
\def\bea{\begin{eqnarray}}
\def\eea{\end{eqnarray}}
\def\obar{\overline}

\def\beqa{\begin{eqnarray}} 
\def\eeqa{\end{eqnarray}} 
\def\beq{\begin{equation}} 
\def\eeq{\end{equation}} 

\def\a{\alpha}          
\def\b{\beta}           
\def\d{\delta}

\def\g{\gamma} 

\def\k{\kappa}
\def\l{\lambda} \def\L{\Lambda}

\def\s{\sigma}

\def\cA{{\cal A}}  \def\cC{{\cal C}}
\def\cD{{\cal D}}  \def\cF{{\cal F}}
 \def\cH{{\cal H}} \def\cI{{\cal I}}
 \def\cK{{\cal K}} 
 \def\cN{{\cal N}} \def\cO{{\cal O}}

 \def\cW{{\cal W}}

\def\mg{\mathfrak{g}}

\newcommand{\R}{\mathds{R}}
\newcommand{\C}{\mathds{C}}

\newcommand{\Z}{\mathds{Z}}

\newcommand{\msu}{\mathfrak{s}\mathfrak{u}}

\newcommand{\mmu}{\mathfrak{u}}

\def\bit{\begin{itemize}}
\def\eit{\end{itemize}}

\def\({\left(}
\def\){\right)}

\def\d{\delta}

 \def\del{\partial}

\newcommand{\tr}{\mbox{tr}}

\def\bcomment#1{}

\newcommand{\co}[2]{[#1,#2]}						
\renewcommand{\a}{\alpha}
\renewcommand{\b}{\beta}
\renewcommand{\d}{\delta}

\renewcommand{\l}{\lambda}
\newcommand{\w}{\omega}

\newcommand{\G}{\Gamma}

\renewcommand{\L}{\Lambda}

\newcommand{\cf}{cf.\ }
\newcommand{\Di}{{\slashed{D}}}

\DeclareMathOperator{\Tr}{Tr}

 \allowdisplaybreaks[3]

\begin{document}

\renewcommand{\title}[1]{\vspace{10mm}\noindent{\Large{\bf
#1}}\vspace{8mm}} \newcommand{\authors}[1]{\noindent{\large
#1}\vspace{5mm}} \newcommand{\address}[1]{{\itshape #1\vspace{2mm}}}


\begin{flushright}
UWThPh-2014-20 
\end{flushright}

\begin{center}

\title{ \Large Self-intersecting fuzzy extra dimensions from squashed coadjoint orbits
in ${\cal N}=4$ SYM and matrix models}

\vskip 3mm

 \authors{Harold C. Steinacker{\footnote{harold.steinacker@univie.ac.at}}, 
 Jochen Zahn{\footnote{jochen.zahn@univie.ac.at}}
 }
 
\vskip 3mm

 \address{ 
{\it Faculty of Physics, University of Vienna\\
Boltzmanngasse 5, A-1090 Vienna, Austria  }  
  }

\vskip 1.4cm

\textbf{Abstract}

\end{center}

We find new vacuum solutions of ${\cal N}=4$ super-Yang-Mills with totally anti-symmetric cubic soft SUSY breaking terms,
or equivalently solutions of the IKKT matrix model of type $\R^4_\theta \times {\cal K}_N$  with flux terms. 
The solutions can be understood in terms of 4- and 6- dimensional fuzzy branes ${\cal K}_N$ in extra dimensions,
describing self-intersecting projections of compact flag manifolds of $SU(3)$. 
The 6-dimensional solutions provide a 6-fold covering  of the internal space near the origin, 
while the 4-dimensional 
branes have  a triple self-intersection spanning all 6 internal directions.
The solutions have lower energy than the trivial vacuum, and
we prove that there are no negative modes. The massless modes are identified explicitly. 
In particular there are chiral fermionic zero modes, linking the coincident 
sheets with opposite flux at the origin. They have a $\Z_3$ family symmetry, originating from the 
Weyl group rotations.

\vskip 2cm

{\em Dedicated to the memory of Bruno Zumino}

\newpage

\tableofcontents

\section{Introduction}
\label{sec:background}

The maximally supersymmetric $\cN=4$ super-Yang-Mills (SYM) theory takes a special role among 
all 4-dimensional field theories. It is arguably the most symmetric 
4-dimensional gauge theory, it is perturbatively finite in the UV 
\cite{Mandelstam:1982cb}, 
and much research effort has been devoted in recent years to study various aspects of this model.
Nevertheless it is far from fully explored. 
In this paper, we find a new class of vacuum solutions in the presence of a cubic soft SUSY breaking 
potential, with remarkable properties.

$\cN=4$ SYM is most naturally obtained by dimensional reduction of 10-dimensional SYM 
to 4 dimensions.
This is also the origin of the IIB or IKKT matrix model, which is obtained by reducing 
 10-dimensional $U(N)$ SYM to zero dimensions \cite{Ishibashi:1996xs}. 
 That model was proposed as a non-perturbative definition
 of IIB string theory, and is expected to contain gravity.
In the large $N$ limit it admits 4-dimensional brane solution $\R^4_\theta$, leading to
4-dimensional noncommutative $\cN=4$ SYM.
Therefore the most obvious approach to 4-dimensional physics from the IIB model 
ought to be via $\cN=4$ SYM. 
However, that theory is usually thought to be  unphysical, due to its 
maximal symmetry, and in particular because of its non-chiral nature.

One way to get more structure from $\cN=4$ SYM is by adding terms to the scalar potential, 
in particular soft SUSY breaking cubic terms. 
It is well-known that such terms lead to non-trivial vacua 
with fuzzy spheres $S^2_N$ in extra dimensions. 
The resulting theory indeed behaves like a 6-dimensional theory with fuzzy extra dimensions, with a 
finite tower of Kaluza-Klein states arising on the fuzzy extra dimensions. 
This is simply a geometric manifestation of the Higgs effect in terms of quantized symplectic spaces, 
and has been studied from various points of view, 
see e.g. \cite{DuboisViolette:1989at,CarowWatamura:1998jn,Myers:1999ps,Iso:2001mg,Polchinski:2000uf,
Berenstein:2002jq,Steinacker:2003sd,Aschieri:2006uw,Andrews:2006aw,Steinacker:2007ay,DelgadilloBlando:2007vx,Chatzistavrakidis:2009ix}.
However, the resulting low-energy theory is still non-chiral.
By choosing suitable cubic terms one may obtain 4-dimensional internal geometries, in particular
$S^2_N \times S^2_N$ \cite{Behr:2005wp}. While this leads to  interesting structures with 
various meta-stable vacua, the resulting low-energy theory is still non-chiral, 
since these 4-dimensional branes have 2 transversal directions in the target space \cite{Chatzistavrakidis:2009ix}. 
There are of course other fuzzy spaces such as $\C P^2_N$ and higher-dimensional 
fuzzy coadjoint orbits of Lie groups $G$, however their standard realization 
requires target spaces with eight dimensions (in the case of $G=SU(3)$) or more, hence they don't fit 
into the $\cN=4$ theory. 
Moreover, their natural embedding in the Lie algebra
$\mg$ of $G$ (in a model with sufficiently many scalar fields) 
would always have  transversal dimensions, and thus lead to a non-chiral low-energy theory.

In order to obtain an effective chiral theory from $\cN=4$ SYM at low energy, 
it seems essential to have branes which ``locally'' span all 6 internal dimensions. 
As shown in \cite{Chatzistavrakidis:2011gs}, chiral fermions indeed appear on 
intersecting branes which span all extra dimensions. 
The doubling problem due to fermions arising on different intersections can be 
avoided\footnote{This issue might also be addressed by warping \cite{Aoki:2014cya}, 
however no such solutions are known at present.} 
by realizing the electroweak Higgs as intrinsic part of the branes,
and backgrounds resembling the standard model can indeed be obtained assuming a suitable 
potential  \cite{Steinacker:2014fja}. 
However, it turns out that such intersecting branes typically lead to some negative modes. Although 
these may be desirable in some sense\footnote{They were associated to a certain singlet Higgs $S$ 
in \cite{Steinacker:2014fja}.}, one would like to have 
a background without any instabilities.

In this paper, we find a new class of classical vacuum solutions of $\cN=4$ SYM
theory on $\R^{3,1}$ with these desired properties, in the presence of suitable cubic terms in the scalar potential. 
These solutions can be interpreted as squashed 4- and 6-dimensional quantized compact spaces or branes
embedded in 6 extra dimensions. They are obtained by a projection of 
quantized coadjoint orbit of $SU(3)$, labeled by dominant integral weights $\mu$ of $\msu(3)$. 
In contrast to previously known solutions,
they have the desired space-filling properties required to get a chiral low-energy theory.
More precisely, the 6-dimensional solutions provide a local cover of $\R^6$ near the origin by 3+3 sheets with opposite
orientation. Similarly, the 4-dimensional solutions  have a triple self-intersection
at the origin, spanning all 6 internal directions.
These features are crucial for the chirality properties of the fermionic zero modes. 
Remarkably, a $\Z_3$ family symmetry
naturally arises on these backgrounds, which originates from the Weyl group
of an underlying $SU(3)$ structure.

After establishing the brane solutions, we study in detail the fluctuations on these backgrounds.
Although we cannot determine the full spectrum, we can show that there are no negative modes,
using a group-theoretical analysis given in appendix \ref{sec:app-pos}.
This is remarkable, since there are typically some negative modes 
for solutions which describe intersecting fuzzy spaces\footnote{We hope to report on these solutions elsewhere.}. 
It turns out that the present backgrounds admit a number of non-trivial zero modes, which are identified explicitly. 
Our analysis is general enough to cover also stacks of various branes of the present type.
The spectrum of the scalar Laplacian governing the 4-dimensional gauge bosons is also determined, which 
only has the expected trivial zero modes corresponding to translations in the internal space.

We then study the fermionic modes on the backgrounds. It turns out that there 
are fermionic zero modes with distinct chirality properties, which we 
work out in detail. 
Since these zero modes link the 3+3 sheets with opposite
orientation, they have fixed chirality determined by the branes on their endpoints.
The point is that the different chiralities are distinguished by their charges w.r.t. the gauge fields
arising on the branes,
which is an essential ingredient for a chiral model.
Analogous chirality properties hold for the zero modes on the 4-dimensional backgrounds.
Assuming that some of the flat deformation modes are switched on, 
this should lead to interesting low-energy effective theories. 
However in the present paper
we restrict ourselves to a technical analysis of these backgrounds.
The analytical results have been verified numerically for small representations.
A more detailed analysis of the possible physical applications is postponed 
to future work, however we point out the similarities
with the backgrounds studied in \cite{Steinacker:2014fja} in an approach to the standard model.

We emphasize that the present solutions make perfect sense at weak coupling in the gauge theory, and 
their interpretation in terms of higher-dimensional geometry does not rely on any
holographic picture. 
It turns out that the classical energy of the new solutions is 
lower than that of the trivial vacuum solution. However, the energy of the well-known
fuzzy sphere solutions is even lower, and the classical ground state is presumably given by a
fuzzy sphere $S^2_N$ with maximal $N$. Therefore the present solutions can only be meta-stable
classically, however quantum corrections might provide sufficient stabilization. 
This should be studied elsewhere.
The solutions also persist in the presence of sufficiently small mass terms, but again we leave a 
detailed analysis for future work.

\section{$\cN=4$ SYM with flux  and Yang-Mills matrix models}

There are two natural starting points for the present paper: either  
maximally supersymmetric $\cN=4$ $U(N)$ SYM with cubic soft SUSY breaking terms, 
or  Yang-Mills matrix models with at least 6 matrices and extra flux terms. 

We start with $\cN=4$ $U(N)$ SYM with extra cubic soft SUSY breaking terms. 
This model is organized most transparently 
in terms of 10 dimensional SYM reduced to 4 dimensions, adding  a cubic potential.
Thus the action is
\begin{align}
 S_{\rm YM}
&=  \int d^4 x \ \tr_N\Big(-\frac 1{4g_{}^2} \cF^{\mu\nu} \cF_{\mu\nu} 
- \frac 12  D^\mu \Phi^a D_\mu \Phi_a + \frac 14 g_{}^2 [\Phi^a,\Phi^b][\Phi_a,\Phi_b] \nn\\
&\qquad \qquad +  \frac i2 f_{abc} \Phi^a \Phi^b \Phi^c  \ 
+ \bar\Psi\g^{\mu}(i\del_\mu + [\cA_\mu,.])\Psi + g_{} \bar\Psi\Gamma^a [\Phi_a,\Psi]\Big) .
\label{N=4SYM}
\end{align}
Here $\cF_{\mu\nu}$ is the field strength, $\Phi^a,\ a \in \cI = \{1,2,4,5,6,7\}$ are 6 scalar 
fields\footnote{The unusual numbering of the indices will become clear below.}, 
$\Psi$ is a matrix-valued Majorana-Weyl spinor of $SO(9,1)$ dimensionally  
reduced to 4-dimensions, and $\Gamma^a, a\in \cI$ arise from the 10-dimensional gamma matrices.
This is discussed in more detail in appendix \ref{sec:clifford}.
 All fields transform in the adjoint of the $U(N)$ gauge symmetry.
The cubic potential specified by  totally anti-symmetric constants $f_{abc}$
will break (most of) the global $SO(6)$ $R$-symmetry, but
respects the internal translational symmetry $\Phi^a \to \Phi^a + c^a \one$ due to the 
anti-symmetry of $f_{abc}$.
These  soft SUSY breaking terms are  expected to preserve the good UV behavior 
of $\cN=4$ SYM \cite{Jack:1999ud}.

\subsection{The 6-dimensional matrix model}

In this paper, we focus on static solutions of the scalar fields of this deformed $\cN=4$ model
corresponding to non-trivial vacua, and the low-energy spectrum arising on such a background.
This is governed by the 6-dimensional Yang-Mills matrix model 
\begin{align}
S_6[X,\Psi] &=  \frac 14\Tr \(\co{X^a}{X^b}\co{X_a}{X_b}\,  
+ 2 i f_{abc} X^a X^b X^c +  4 \obar\Psi \Gamma_a[X^a,\Psi] \)   
\label{V-general-flux}
\end{align}
for $a,b,c \in \cI$, 
with totally anti-symmetric constants $f_{abc}$. 
We  write $X^a$ instead of $\Phi^a$
to emphasize their geometric significance as quantized embedding coordinates 
in the internal $\R^6$. 
Here $X^a$ are hermitian $N\times N$ matrices
 i.e. operators acting on the  Hilbert space $\mathcal{H}=\C^N$, and
the coupling constant $g$ is absorbed in $X^a$ resp. $\Psi$.
This action is invariant under gauge transformations $X^a \to U X^a U^{-1}$ for $U\in U(\cH)$, 
and translations $X^a \to X^a + c^a \one$
(up to  traces of commutators), due to the anti-symmetry of $f_{abc}$.
The cubic potential breaks some or all of the global $SO(6)$ symmetry.

The spinors arise from a matrix-valued Majorana-Weyl (MW) spinor of $SO(9,1)$, with Clifford generators $\G_a$. 
Hence we view $\Psi \in \C^4 \otimes \C^8 \otimes Mat(N,\C)$ as Dirac spinor on $\R^6$ taking values
in $\C^4\otimes Mat(N,\C)$, subject to the MW constraints. 
Factorizing  the 10-dimensional chirality and charge conjugation operators
(see appendix \ref{sec:clifford}), the 10-dimensional MW constraints can be written  as 
\begin{align}
 \G^{(6)} \Psi &= \g_5 \Psi,  \nn\\
 C^{(6)}  \Psi^* &= - C^{(4)} \G^0 \Psi \ .
 \label{MW-decomp}
\end{align}
This means that the internal chirality of the spinor determines the 
4-dimensional chirality, and similarly the (anti-linear) charge conjugation of the internal 
spinor is equivalent to the (linear) charge conjugation operator
$- C^{(4)} \G^0$
acting on the 4-dimensional spinor. Here $*$ denotes conjugation of the $N\times N$ matrices.

This is the model we will study in the remainder of this paper.
Similar models have been studied in considerable detail, see e.g. 
\cite{Alekseev:2000fd,CarowWatamura:1998jn,Iso:2001mg,Steinacker:2003sd,Andrews:2006aw,Steinacker:2007ay,
DelgadilloBlando:2007vx}
just to name a few pertinent works. 
We will find new solutions of this 6-dimensional matrix model
for a suitable choice of $f_{abc}$ \eq{cubic-complex}, 
which leads to rich and very interesting internal geometries.
As discussed below, our results  also apply to the IKKT model reduced to
noncommutative  $\cN=4$ SYM on commutative $\R^4_\theta$, and are also relevant to 
the BFSS model on 3-dimensional space-times.

Finally we note that
the constant 4-dimensional gauge fields $A_\mu$ are 
governed by the action $\frac 12\Tr [A_\mu,X^a][A^\mu,X_a] = -\frac 12\Tr A_\mu,[X_a,[X^a,A^\mu]]$. 
On an irreducible brane solution,
this will give a mass to the $A_\mu$ via the Higgs effect for all but the 
trivial modes of $A_\mu$, as discussed in Section \ref{sec:scalar-Laplace}. 
Hence we can safely set $A_\mu = 0$ for the background.

\paragraph{Fluctuations and equations of motion.}

Now consider fluctuations around any given background $\bar X^a$ of this model,
\begin{align}
 X^a = \bar X^a + \cA^a
\end{align}
where $\cA^a$ are arbitrary matrices\footnote{On an irreducible brane background, they can 
be written as functions on the brane, $\cA^a(\bar X)$. On a stack of $N$ coinciding branes, 
they can be written as $\mmu(N)$-valued functions on the brane.}.
Then the action \eq{V-general-flux} expanded up to second order in $\cA^a$ is\footnote{We do not distinguish 
upper and lower indices of the $X^a$ matrices.}
\begin{align}
 S_6[X] &= S_6[\bar X]  + \frac 12\Tr \Big(2[\cA^a,\bar X^b][\bar X_a,\bar X_b] 
 +  3i  f_{abc} \bar X^a\bar X^b \cA^c \nn\\
 &\qquad + [\bar X^a,\cA^b][\bar X_a,\cA_b] +  [\bar X^a,\cA^b][\cA_a,\bar X_b] 
  + [\bar X^a,\bar X^b][\cA_a,\cA_b] 
  + 3i f_{abc} \cA^a\cA^b \bar X^c   \Big) \nn\\
 &= S_6[\bar X]  + \frac 12\Tr \Big(\cA^a( -2 \Box \bar X_a + 3 i f_{abc} \bar X^b\bar X^c)  \nn\\
 &\qquad  -\cA_a \Box \cA^a + 2 [\cA_a,\cA_b] \big([\bar X^a,\bar X^b]+ \frac{3}4 i f_{abc} \bar X^c \big) + f^2  \Big)
\label{eff-S-expand}
\end{align}
 dropping fermionic terms and using
\begin{align}
 \Tr [\bar X^a,\cA^b][\cA_a,\bar X_b]  
  &=  \Tr( [\cA_b,\cA_a][\bar X^b,\bar X^a] + [\cA^b,\bar X_b],[\bar X^a,\cA_a]) 
\end{align}
with the Jacobi identity. Here  
 \begin{align}
 \Box \cA^a \equiv [\bar X^b,[\bar X_b,\cA^a]] \ .
\end{align}
We observe that
\begin{align}
  f = i[\bar X_a,\cA^a] 
 \label{gauge-fixing-function}
\end{align}
can be viewed as gauge fixing function, since it
transforms as 
\begin{align}
 f \to f + \Box \L
\end{align}
under infinitesimal gauge transformations $U=e^{i\L}$. We can thus choose the gauge such that $f=0$. 
This term disappears using either the Faddeev-Popov
procedure or BRST procedure in the matrix model, which we will exploit in 
section \ref{sec:fluctuations}.

The equations of motion for the background are 
\begin{align}
 \Box X^a  = \frac 32 i f_{abc}  X^b X^c \ 
\end{align}
(dropping the bar from now on).
It is easy to check that configurations with
\begin{align}
 [ X^a, X^b]+ \frac{3}4 i f_{abc} X^c = 0 
 \label{fuzzy-solution-basic}
\end{align}
are always solutions, since then
\begin{align}
  \Box X^b &= [X_a,[X^a,X^b]] =  -i \frac{3}4 f_{abc} [X_a, X^c]
  = \frac{3}2 i f_{bac} X_a X^c .
\end{align}
The fluctuations $\cA^a$ are governed by the linearized ``vector'' equations of motion
\begin{align}
\Big(\Box\d^a_b + 2 [([X^a, X^b] +\frac{3}4 i f_{abc} X^c),. \, ]  
 - [X^a,[X^b,.]]\Big) \cA_b = 0 \ .
\label{vector-eom}
 \end{align}

\paragraph{IIB matrix model reduced to $\R^4_\theta$.}

The 6-dimensional matrix model \eq{V-general-flux} also governs 
the internal sector of the IIB or IKKT matrix model on a stack of 
 $N$ coinciding $\R^4_\theta$ branes. 
We recall that $\R^4_\theta$ is defined as irreducible representation of 
$[\bar X^\mu, \bar X^\nu] = i \theta^{\mu\nu} \one$, 
where $\theta^{\mu\nu}$ has rank 4. This is a brane solution of the IKKT model.
Fluctuations  of the matrices around on a stack of $N$ such coincident $\R^4_\theta$ branes
are governed by 
maximally supersymmetric {\em non-commutative} $\cN=4$ $\ U(N)$ Super-Yang-Mills theory
 on $\R^4_\theta$.
Interpreting the fluctuations $\cA^a$ as 
$\mmu(N)$-valued functions in $\R^4_\theta$, 
the matrix model reduces to (cf. \cite{Aoki:1999vr,Steinacker:2010rh})
\begin{align}
 S_{\rm YM} 
&=  \int d^4 x\, \sqrt{G} \ \tr_N\Big(-\frac 1{4g_{}^2} (\cF \cF)_G 
- \frac 12 G^{\mu\nu} D_\mu \Phi^a D_\nu \Phi_a + \frac 14 g_{}^2 [\Phi^a,\Phi^b][\Phi_a,\Phi_b] \nn\\
&\qquad \qquad   + \bar\Psi\tilde\g^{\mu}(i\del_\mu + [\cA_\mu,.])\Psi + g_{} \bar\Psi\Gamma^a [\Phi_a,\Psi]\Big) 
\label{S-YM-eff}
\end{align}
where 
\begin{align}
 X^\mu &= \bar X^\mu + \theta^{\mu\nu} \cA_\nu ,  \qquad \mu,\nu = 0,...,3 \nn\\ 
G^{\mu\nu} &= \rho \theta^{\mu\nu'}\theta^{\nu\nu'} \eta_{\mu'\nu'} , 
\qquad \tilde\g^{\mu} = \rho^{1/2}\, \theta^{\nu\mu} \gamma_\nu   \nn\\
   \rho &= \sqrt{|\theta^{-1}|}, 
\label{field-identification} 
\end{align} 
and  $\cF_{\mu\nu} = \del_\mu \cA_\nu - \del_\nu\cA_\mu + [\cA_\mu,\cA_\nu]$ is the $\mmu(N)$ field strength.
The scalar fields $\Phi^a$ and the spinors $\Psi$ are functions on 
noncommutative $\R^4_\theta$ taking values in $\mmu(N)$, with analogous constraints as
in the commutative case.
Although the action \eq{S-YM-eff} is written in a way that looks like the standard $\cN=4$ SYM, 
it is in fact noncommutative $\cN=4$ SYM on $\R^4_\theta$. 
We can add again cubic terms corresponding to soft SUSY breaking terms, 
which are expected to preserve the good UV behavior 
of the IKKT model on a stack of 4-dimensional branes \cite{Jack:2001cr}. 
We then recover the action \eq{V-general-flux} which governs its internal sector.

\section{Fuzzy sphere solutions}
\label{sec:fuzzy-sphere}

We first recall the  fuzzy sphere solution of this model, 
assuming a simple flux term $f_{abc} = f \varepsilon^{123}_{abc}$ for simplicity.
Using \eq{fuzzy-solution-basic}, the equations of motion are solved by
\begin{align}
 X_a = -\frac 38\ f\  T_a 
 \label{S2-solution}
\end{align}
where the $T_a, \  a = 1,2,3$ is any  representation of $\msu(2)$,
\begin{align}
 [T_a,T_b] &= 2i \varepsilon_{abc} T_c \ .
\end{align}
Taking $T_a$ to be the $N$-dimensional irreducible representation of  $\msu(2)$,
this is the (rescaled) fuzzy sphere, with \cite{Madore:1991bw,hoppe}
\begin{align}
  T_1^2+T_2^2+T_3^2 &= N^2-1 =: R^2 \ .
\end{align}
This is a quantization of the classical 
sphere with quantized embedding functions $T_a \sim t_a$
and Poisson structure $\{t_a,t_b\} = 2 \varepsilon_{abc} t_c$,
carrying $N$ units of flux. 
For these ``brane'' solutions, the equations of motion \eq{vector-eom} for the fluctuations 
$X_a = \bar X_a + \cA_a$ 
simplify as
\begin{align}
 \Box\cA_a = 0
 \label{fluct-sphere}
\end{align}
assuming the gauge $[\bar X^a,\cA_a] =0$.
This means that there are massless deformations of the branes (corresponding to translations), 
but no negative modes.

\subsection{The squashed fuzzy sphere}
\label{sec:squashed-sphere}

As a warm-up for the new solutions, we discuss the squashed fuzzy sphere 
obtained by projecting $S^2_N$ onto the equatorial plane. 
This corresponds to a 2-sheeted fuzzy disk glued together at the boundary, illustrating the 
idea of gluing branes in \cite{Steinacker:2014fja}.
It also illustrates the strategy of obtaining matrix model solutions 
based on Lie algebras  discussed in \cite{Kim:2012mw}.

We start with $S^2_N$ defined by three matrices $T_a$ as above. 
Now consider the background defined by the {\em two} matrices
\begin{align}
 X_1 = T_1, \quad X_2 = T_2 \ . 
\end{align}
This describes a 2-sheeted fuzzy disc, given by the projection of $S^2_N$ along the 
Cartan generator $X^3$. Note that in the semi-classical limit, 
the two sheets have opposite orientation as defined by their local
Poisson structure\footnote{Note that this differs from the Poisson structure on the 
fuzzy disk introduced in \cite{Lizzi:2003ru}.}
$\{x^1,x^2\} = \theta^{12}(x) = \pm 2R\sqrt{1-\frac{x_1^2+x_2^2}{R^2}}$, and are glued together at the boundary. 
This is a solution of the matrix equations 
\begin{align}
 \Box_X X_i &= 4 X_i, \qquad i = 1,2
 \end{align}
 where
 \begin{align}
 \Box_X &=  [X_1,[X_1,.]] + [X_2,[X_2,.]] =  2  ([X^+,[X^-,.]] + [X^-,[X^+,.]])  \nn\\
  \Box &=  [T_1,[T_1,.]] + [T_2,[T_2,.]]  + [T_3,[T_3,.]] 
\end{align}
and
\begin{align}
 X^\pm &= \frac 12(X_1\pm i X_2) .
\end{align}
Now consider the Dirac operator on such a squashed sphere:
\begin{align}
\Di_s \Psi &=  \sum_{a=1,2} \s_a [X^a,.]  =  2\Big(\s^- [X^+,.] + \s^+ [X^-,.]\Big) 
\end{align}
where $\s^- = \frac 12(\s_1-i\s_2), \ \s^+ = \frac 12(\s_1+i\s_2)$.
It anti-commutes with the chirality operator $\s_3$.
To compute its spectrum, we consider
\begin{align}
\Di_s^2 \Psi 
 &= \Box - [T_3,[T_3,.]] - 2 \sigma_3[T_3,.] \nn\\
  &= \Box - ([T_3,.] + \s_3)^2 + 1 .
 \label{Dirac-ladder-disc}
\end{align}
Decomposing the space of functions 
$Mat(N,\C) = \C^N \otimes {\C^N}^* = \oplus_{l=0}^{N-1} \C^{2l+1}$ with basis $|l,m_l\rangle$
in quantum mechanics notation
and passing to the 
total angular momentum basis of $\C^2\otimes \C^{2l+1}$ labeled by $j,l,m_j$, we can read off the eigenvalues
\begin{align}
 E_{jlm_j}^2 &= 4 l(l+1) - 4 m_j^2 + 1  = 4(l+\frac 12-m_j)(l+\frac 12+m_j)
\end{align}
where $m_j = m+s$, and $s$ is the eigenvalue of $\frac 12 \s_3$.
Hence for each $l\in \{0,1,2,...,  N-1\}$, there is pair of zero modes with extremal 
weights $m_j=\pm(l+\frac 12)$,  which in the $\C^2\otimes Mat(N,\C)$ basis take the form
\begin{align} 
 \Psi_+ = |\uparrow\rangle |l,l\rangle , \qquad \Psi_- = |\downarrow\rangle |l,-l \rangle
\end{align} 
since $X^+ |\uparrow\rangle |l,l\rangle = 0 = \s^+|\uparrow\rangle |l,l\rangle $ etc. 
Thus the fermionic zero modes correspond to the
extremal weights in the angular momentum decomposition of $Mat(N,\C)$. 
For example, the maximal weight state in $Mat(N,\C)\cong \C^N \otimes {\C^N}^*$
can be written as
\[
 |N-1,N-1\rangle = |\tfrac 12(N-1)\rangle\langle -\tfrac 12(N-1)|,
\]
where $|\pm\frac 12(N-1)\rangle$ 
can be interpreted as coherent states\footnote{The concept of coherent states on fuzzy 
spaces will be discussed in more detail in section \ref{sec:geometry}.} located at the origin of the two sheets.
Therefore these zero modes can be interpreted as fermionic strings, linking the two opposite fuzzy sheets 
at or near the origin.
Moreover, the zero modes with opposite chirality are in fact, up to a phase, charge conjugates of each other, $C \Psi_+^* = i \Psi_-$ with $C = \s_2$ the charge conjugation matrix, \cf Appendix~\ref{sec:clifford}. 

Since these fermionic strings link  (sheets of) 
noncommutative branes with opposite flux defined by the symplectic structure $\omega$, 
they behave like charged particles on a single (noncommutative) disk
with flux $(2\pi)^{-1}\int \omega = N$, in analogy to the Landau problem. 
This leads to $N$ zero modes for each charge i.e. orientation of the string,
and the chirality of these zero modes is determined by their charge.
To make this more explicit, consider the $U(1)$ symmetry generated by $T_3$, 
 corresponding to rotations  in the $1-2$ plane. 
Then although  the chiral zero modes arise in pairs with opposite chirality
as the underlying  model is non-chiral,
the different chiralities are nevertheless distinguished by the charges w.r.t.  $T_3$. 
Of course that symmetry is broken by the 
squashed sphere background (so that the corresponding gauge fields 
$A_\mu^{(3)}(x)$ would be massive; this will be manifest in the background solutions below),
nevertheless this is a signature of a chiral model, 
such as the standard model in the broken phase.


This example illustrates the importance of (locally) space-filling branes for 
chiral fermions\footnote{Such (locally) space-filling branes were  missing  in previous attempts 
such as \cite{Steinacker:2007ay}.} in extra dimensions.
However, the squashed fuzzy sphere is not a solution. 
In the remainder of this paper, we will find 4-and 6-dimensional fuzzy brane {\em solutions} 
with analogous properties.

%

\section{Coadjoint orbits of $SU(3)$ and projections}

We consider the Lie algebra $\mg = su(3)$ with 8 generators $T_a,\ a=1,...,8$, canonically normalized 
with respect to the Killing form
\begin{align}
\k_{ab} = (T_a,T_b) = 2 \d_{ab}
\end{align}
and structure constants
\begin{align}
 [T_a,T_b] = ic_{abc} T_c
 \label{structure-const}
\end{align}
The Cartan subalgebra $\mg_0 \subset \mg$ is spanned by the  
two orthogonal Cartan generators $H_3 = T_3$ and $H_8 = T_8$.
The fundamental representation $\cH_{(1,0)}$ of $su(3)$ corresponds to the  Gell-Mann matrices 
$\l_a = \pi_{(1,0)}(T_a)$, 
\begin{align}
 [\l_a,\l_b] = ic_{abc} \l_c
\end{align}
canonically normalized as
\begin{align}
 \tr(\l_a\l_b) = 2\d_{ab} .
\end{align}
The structure constants are explicitly
\begin{align}
 {c_{12}}^3 &= 2, \; {c_{14}}^7 = 1,\; {c_{15}}^6 = - 1,\; 
  {c_{24}}^6 = 1, \; {c_{25}}^7 = 1, \nn\\
{c_{34}}^5 &= 1, \; {c_{36}}^7 = - 1,\; 
{c_{45}}^8 = \sqrt{3}, \; {c_{67}}^8 = \sqrt{3},
\label{f-explicit}
\end{align}
while all others are zero.

\subsection{Classical geometry}

Coadjoint orbits of $SU(3)$ are defined as
\begin{align}
 \cC[\mu] = \{g^{-1} \mu g; \ g \in SU(3) \} \cong SU(3)/\cK
 \label{class-orbit}
\end{align}
where $\cK$ is the stabilizer group of some weight 
(or charge) $\mu \in \mg_0^*$,
or equivalently of $T_\mu \in \mg_0$ in the adjoint. In general, such coadjoint orbits are known as flag manifolds.
Here $T_\mu$ can have two or three different eigenvalues, leading to
4- or 6-dimensional flag manifolds $\cC^4[\mu] \cong \C P^2$ resp $\cC^6[\mu]$.
If $T_\mu$ has two almost degenerate eigenvalues,
it is natural to
consider $\cC^6[\mu]$ as $S^2$ bundle over $\C P^2$, noting that 
$\cK_6 = U(1)\times U(1) \subset \cK_4 = SU(2)\times U(1)$. 
The corresponding map
\begin{align}
 SU(3)/\cK_6  \to  SU(3)/\cK_4  
\end{align}
defines the $S^2$ foliation, since ${U(1)}\backslash SU(2)\cong S^2$.
We observe that $\cC[\mu]$ is 6-dimensional if the Weyl group $\cW$ acts
freely on $\mu$, and it is 4-dimensional if $\mu$ has a non-trivial stabilizer in $\cW$.
 
The coadjoint orbit $\cC[\mu]$ is regularly embedded in $\R^8 \cong \mg^*$, with coordinates $x^a,\ a = 1,...,8$. 
The 6-dimensional  orbits are characterized by 
the two Casimirs
\begin{align}
 C_2 = \sum_{a=1}^8\, x^a x^a , \qquad C_3  =  \sum_{a,b,c=1}^8\, d_{abc} x^a x^b x^c .
\label{C2-C3}
 \end{align}
The latter is given explicitly by
\begin{align}
C_3 &= \sqrt{3} x^8(-2 x^1 x^1-2 x^2 x^2 -2 x^3 x^3 + x^4 x^4  + x^5 x^5  + x^6 x^6 + x^7 x^7  + 2x^8 x^8 ) \nn\\
& \quad + 3 x^3 (- x^4 x^4  - x^5 x^5 + x^6 x^6 + x^7 x^7) 
 + 6 x^6(-x^1 x^4 - x^2 x^5)  + 6 x^7(x^2 x^4 - x^1 x^5)  \nn
\end{align}
using the explicit $d_{abc}$ constants
\begin{align}
{d_{11}}^8 &= -2/\sqrt{3}, \;  
{d_{22}}^8 = -2/\sqrt{3},    \;  
{d_{33}}^8 = -2/\sqrt{3}, \; \;
 {d_{44}}^8 = 1/\sqrt{3}, \nn\\
 {d_{55}}^8 &= 1/\sqrt{3},\;  
 {d_{66}}^8 = 1/\sqrt{3}, \; 
 {d_{77}}^8 = 1/\sqrt{3}, \; 
 {d_{88}}^8 = 2/\sqrt{3},\nn\\
 {d_{34}}^4 &= -1,\; 
  {d_{35}}^5 = -1,\; 
  {d_{36}}^6 = 1,\; 
 {d_{14}}^6 = -1,\;
 {d_{25}}^6 = -1,\;
 {d_{24}}^7 = 1,\;
 {d_{15}}^7 = -1,\; 
  {d_{37}}^7 = 1 .
\nn  
\end{align}
We can generically solve these two equations for the ``Cartan coordinates'' $x^3$ and $x^8$,
 with 6 different branches.
In contrast, the 4-dimensional orbits are characterized by the constraints \cite{Alexanian:2001qj,Grosse:2004wm}
\begin{align}
  \sum_{a=1}^8\, x^a x^a = C_2, \qquad 
 \sum_{a,b=1}^8 d_{abc} x^a x^b  = \sqrt{\frac{4C_2}3}\ x^c \ . 
 \label{CP2-constraints}
\end{align}
Now consider the  projection $\Pi$ of $\cC[\mu]$ on $\R^6$, defined as
\begin{align}
\cC[\mu] \ \hookrightarrow \R^8 \  & \ \stackrel{\Pi}{\to} \ \R^6   \nn\\
   (x^a)_{a=1,...,8} & \ \mapsto \ (x^a)_{a=1,2,4,5,6,7}
 \label{projection}
\end{align}
This gives a squashed 4- or 6-dimensional orbit with singular self-intersecting embedding in $\R^6$.
Due to the multiple covering, the algebra generated by these 6 functions $x^a, \ a\in\cI = \{1,2,4,5,6,7\}$ 
on $\cC[\mu]$  cannot distinguish the different branches, and defines a proper subalgebra of
the full algebra of functions on $\cC[\mu]$.
In contrast, we will see  that the corresponding  algebra of functions
on the quantized orbits is the same before and after the reduction.

\subsection{The fuzzy coadjoint orbits $\C P^2_N$ and $\cC^6_N[\mu]$}

Consider a classical coadjoint orbit  defined by $\mu$ as above. We can assume 
that $\mu$ is in the fundamental Weyl chamber. 
For simplicity, we assume that $\mu = n_1\L_1+n_2\L_2 \equiv (n_1,n_2)$ is a dominant 
integral weight\footnote{In general, the ray $\R\mu$ can be approximated by a sequence of 
dominant integral weights $\mu_N$.}.
Let $\cH_{\mu}$ be the corresponding highest weight irreducible representation.
Then the fuzzy space $\cC_N[\mu]$ corresponding\footnote{The integer $N$ indicates that the space is 
quantized in terms of some finite-dimensional representation associated with $N$.
The precise characterization is done in terms of the highest weight $\mu$.} 
to the above coadjoint orbits 
is defined in terms of 8 hermitian matrices 
\begin{align}
 X^a = \pi_{\mu}(T^a)
\end{align}
where $T^a$ denotes\footnote{We do not distinguish 
between upper and lower indices here.} 
the generators of $SU(3)$, and $\pi_{\mu}$  the representation on $\cH_\mu$.
They generate the simple matrix algebra
\begin{align}
 \cA = End(\cH_\mu) .
\end{align}
Under the adjoint action of  $SU(3)$,
this  decomposes into the irreducible representations 
\begin{align}
  \cA = End(\cH_\mu) = \oplus_\L n_\L \cH_\L .
  \label{harmonic-decomp}
\end{align}
One can show  (see e.g. \cite{Pawelczyk:2002kd}) that the  multiplicities $n_\L$
coincide with the corresponding decomposition of the classical functions on $\cC[\mu]$
into harmonics, for $\L$ below some cutoff. 
Therefore $\cA = End(\cH_\mu)$
can be interpreted as quantized algebra of functions on $\cC[\mu]$.
The matrices $X^a$ can be interpreted as quantized embedding functions 
\begin{align}
 X^a  \ \sim x^a: \quad \cC[\mu] \ \hookrightarrow \ \R^8 .
\end{align}
They satisfy the commutation relations
\begin{align}
 [X^a,X^b] = i c_{abc} X^c
\end{align}
which is a quantization of the canonical Kirillov-Kostant symplectic form on $\cC[\mu]$, 
as well as  constraint equations analogous to \eq{C2-C3}, \eq{CP2-constraints} 
with specific  constants 
depending on the representation $\cH_\mu$.

The simplest example of such a fuzzy coadjoint orbit is the 
fuzzy sphere $S^2_N$, which is the quantized coadjoint orbit $S^2$ of $SU(2)$
defined by the $N$-dimensional irreducible representation of $SU(2)$.

An interesting class of fuzzy flag manifold associated to $SU(3)$ is defined by $\mu = (N,0)$ 
resp. $\mu = (0,N)$. This leads to quantized fuzzy $\C P^2_N$ 
\cite{Alexanian:2001qj,Grosse:1999ci,Grosse:2004wm}, which is
a 4-dimensional fuzzy flag manifold with functions decomposing into harmonics as follows
\begin{align}
 \cA = End(\cH_\mu) = \bigoplus\limits_{n=0}^N \ \cH_{(n,n)} .
  \label{harmonic-decomp-CP2}
\end{align}
Another interesting class are the fuzzy spaces with $\mu = (N,1)$ 
resp. $\mu=(1,N)$, with harmonic decomposition
\begin{align}
 \cA = End(\cH_\mu) = (N+1,N+1)  \oplus (0,0) \bigoplus\limits_{n=1}^N \Big(2 (n,n)
  \oplus  (n-1,n+2) \oplus (n+2,n-1) \Big) .
  \label{harmonic-decomp-CP2-S2}
\end{align}
For example, the space of functions on the fuzzy space defined by $\mu = (1,1)$
decomposes into
\begin{align}
 End(\cH_{(1,1)}) = (1,1)\otimes (1,1) = (2,2) \oplus (3,0)\oplus (0,3) \oplus 2(1,1) + (0,0) .
\end{align}

\subsection{Squashed fuzzy  orbits of $SU(3)$}

The Lie algebra  $\mg =\msu(3)$ has two simple roots $\a_{1,2}$ and one composite root 
$\a_3 = \a_1+\a_2$.
We single out the corresponding  root generators, denoted as
\begin{align}
 X_1^\pm &= \frac 12(T_4\pm i T_5) , \nn\\
 X_2^\pm &= \frac 12(T_6\mp i T_7), \nn\\ 
 X_3^\pm &= \frac 12(T_1\pm i T_2) 
 = \pm [X_1^\pm,X_2^\pm]
 \label{X-T-definition}
\end{align}
or simply set
\begin{align}
 X_a = T_a, \quad a\in  \cI := \{1,2,4,5,6,7\} \ .
\end{align}
For any irrep $\cH_\mu$ of $\msu(3)$, this provides us with 6 hermitian matrices 
$\pi_\mu(X^a) \equiv X^a_{(\mu)}$ for $a\in\cI$ (we will often drop the subscript $\mu$). 
This defines a squashed fuzzy coadjoint orbit obtained from $\cC_N[\mu]$, 
interpreted as noncommutative brane with singular embedding in $\R^6$.
Note that the Cartan generators are {\em not}
included in the $X_a$, which corresponds to the projection $\Pi$ \eq{projection}.
This is the background under consideration here.

Now recall the canonical isomorphism
\begin{align}
 \mg_0^*  \  &\to \ \mg_0   \nn\\
  \a  &\mapsto  H^\a
\label{isomorphism}
\end{align}
defined by the Killing form.
These Cartan generators satisfy
\begin{align}
 H^\a|\mu\rangle = (\a,\mu) |\mu\rangle
\end{align}
for any weight vector $|\mu\rangle$ in a representation.
In particular, the usual Cartan generators correspond to
\begin{align}
 T_3 = H^{\a_3} = H_3   , \quad T_8 = H^{\a_8} = H_8
\end{align}
where $\a_3, \a_8$ form an orthogonal basis of $\mg_0^* \cong \R^2$ normalized as
\begin{align}
 (\a_3,\a_3) = 2 = (\a_8,\a_8) .
\end{align}
A general element $H \in \mg_0$ satisfies
\begin{align}
 [H,X_i^\pm] = \pm\a_i(H) X_i^\pm \ .
\end{align}
Explicitly, the commutators of the generators $X^a, \ a\in \cI$ of squashed $\cC_N[\mu]$
are given by
\begin{align}
 [X_i^+,X_i^-] &=  H^{\a_i}, \qquad i = 1,2,3  \nn\\
 [X_1^+,X_2^+] &=  X_3^+  \nn\\
 [X_1^+,X_3^-] &= - X_2^-  \nn\\
 [X_2^+,X_3^-] &=  X_1^-  \nn\\
 [X_1^+,X_2^-] = [X_2^+,X_3^+] = [X_1^+,X_3^+] & = 0
 \end{align}
as well as the  conjugate relations. 
Note that these relations close up to elements of the Cartan subalgebra.
Nevertheless, the  $X^a$ for  $a\in\cI$ generate the full matrix algebra of functions 
on fuzzy $\cC_N[\mu]$, in contrast to the classical case. This implies that the 
Kaluza-Klein modes on squashed fuzzy $\cC_N[\mu]$ are in one-to-one correspondence to those on 
ordinary fuzzy $\cC_N[\mu]$, however the effective geometry and the 
spectrum of the relevant Laplacian will be different. In particular, we will be able to distinguish 
different branches of squashed fuzzy $\cC_N[\mu]$.

Now consider the corresponding matrix Laplacian on squashed $\cC_N[\mu]$
\begin{align}
 \Box_{X} &:= \sum_{a \in\cI} [X_a,[X_a,.]] \nn\\
  &= 2 \sum_{i=1}^3  ([X_i^+,[X_i^-,.]] + [X_i^-,[X_i^+,.]])  .
\end{align}
It preserves the $\msu(3)$ charges, so that the ladder operators $X_i^\pm$ must be eigenvectors.
One finds
\begin{align}
 \Box_{X} X_a &= 8 X_a , \qquad a \in \cI .
 \label{Box-X}
\end{align}
This is easily seen by direct verification, e.g.
 $\Box_X X_1^+ = 2 (2 +2) X_1^+  = 8 X_1^+$.
Now consider the term $g^{abc} [X^b, X^c]$, where 
\begin{align}
 i g_{abc} &= \frac 1{2c_N}\tr(X_a [X_b,X_c]) = \frac 1{2}\tr(\l_a [\l_b,\l_c])   \nn\\
\label{eq:def_g}
 \k_{ab} &= \frac 1{c_N} \tr(X^a X^b) = 2 \d_{ab}
\end{align}
are the totally anti-symmetric structure constants with indices reduced to the index set $\cI$. 
By this definition and recalling \eq{structure-const}, it follows that
\begin{align}
 [X_a,X_b] - i g_{abc} X_c  &= i \sum_{i=3,8} c_{abi} H_i \nn\\
  &= \frac 12 \big(\tr(\l_a [\l_b,\l_3]) H_{3} + \tr(\l_a [\l_b,\l_8]) H_8 \big) 
 \label{epsilon-Cartan-rel}
 \end{align}
where 
\begin{align}
 i c_{abi} &= \frac 1{2}\tr(\l_a [\l_b,\l_{i}]) \ .
\end{align}
Again, $i g_{acb} [X_a,X_c]$ must be proportional to $X_b$  since the 
$\msu(3)$ charges are preserved, and  by direct verification one finds
\begin{align}
  i g_{acb} [X_a,X_c] = -4 X_b  
 \label{gXX-relation}
\end{align}
(this follows e.g. from
\begin{align}
  \sum_{a,c}g_{acb}g_{acb'} & = 4 \d_{bb'}, &
 \sum_{a,c} g_{acb} c_{aci} & = 0,
  \label{gg-sum}
\end{align}
which is easily verified).
Combining \eq{Box-X} and \eq{gXX-relation}, 
we obtain  
\begin{align}\fbox{$ \ 
 \Box_X X^b = -2 i g_{acb} [X_a,X_c] \ .
 $}
\end{align}
This means that the $X_a, \ a\in \cI$ are a 
solution of the matrix model \eq{V-general-flux} with the totally anti-symmetric flux terms
\begin{align}
\label{eq:f_abc}
 f^{abc} = -\frac 83 g^{abc} \ .
\end{align}
In terms of  the complex combinations $X_i^\pm$ \eq{X-T-definition}, these cubic terms in the action 
can be succinctly written as\footnote{This is equivalent to the form $32\Tr\big( [X_1^+,X_2^+] X_3^+ + h.c.\big)$
upon a substitution $X_2 \to -X_2, \ \Gamma^2 \to -\Gamma^2$.}
\begin{align}\fbox{$ \
 \Tr \big(i f_{abc} X^a X^b X^c\big) 
  = 32\Tr\big( [X_1^+,X_2^+] X_3^- + h.c.\big) \ .
 $}
\label{cubic-complex}
\end{align}
We note that $f_{abc}$ is a combination of selfdual and anti-selfdual 3-forms on $\R^6$,
The potential is invariant under the Weyl group of $SU(3)$,
which acts on the fields as
\begin{align}
 S_3\circ \begin{pmatrix}
           X_1^+\\ X_2^+\\ X_3^-
          \end{pmatrix}
         = \begin{pmatrix}
            X_2^- \\ X_1^- \\ X_3^+
           \end{pmatrix}
\end{align}
and analogously for the other reflections $S_1, S_2$.
Such cubic terms are known to arise from string theory compactifications on Calabi-Yau manifolds with fluxes,
cf. \cite{Camara:2003ku,Grana:2003ek}. 
There is a lot of literature on similar actions, and it is well-known that 
fuzzy spheres arise as solutions. However, our cubic terms are distinct from those e.g. in 
the $\cN=1^*$ action \cite{Polchinski:2000uf,Andrews:2006aw}, which admits
fuzzy spheres as exact supersymmetric ground states in the Higgs phase. 
Supersymmetry is broken explicitly in our translationally invariant action,
which besides the fuzzy sphere solutions also admits squashed higher-dimensional 
fuzzy spaces $\cC_N[\mu]$ as solutions. 
Moreover, we will show that there are  no 
negative modes on these backgrounds\footnote{In fact the above action has other 
remarkable solutions similar to those considered in \cite{Steinacker:2014fja}, 
which however tend to have some negative modes.}.

\subsection{Fluctuation modes}
\label{sec:fluctuations}

Now consider the quadratic fluctuations around such a background,
which give rise to  scalar or Higgs fields on $\R^4$.
Using    \eq{eff-S-expand}, they are governed by the potential
\begin{align}
 V[\cA] = - S[\cA] &= \tr\Big(\cA_a  \Big(\frac 12\Box_X\d^a_b +  [([X_a, X_b] - 2i g_{abc} X_c),. \, ] 
  \Big) \cA_b - \frac 12 f^2\Big) , \nn\\
  f &= i [X_a,\cA_a]
 \label{quad-action-fluct-nogf}
\end{align}
expanded up to quadratic order in $\cA_a$. This leads to 
the vector Laplace operator \eq{vector-eom}
\begin{align}
(\Box_V \cA)_a = 
\Big(\Box_X\d^a_b + 2 [([X_a, X_b] - 2i g_{abc} X_c),. \, ] - [X^a,[X^b,.]]  \Big) \cA_b \ .
\label{vector-laplace}
 \end{align}
This operator generically has $\dim(\cH)^2-1$ zero modes due to the $U(\cH)$ gauge symmetry.
However these are unphysical, and we are only interested in the physical spectrum after 
dropping these pure gauge modes.
In the matrix model without $\R^4_\theta$ background, this is achieved by adding a gauge fixing term
to the action. On a background with $\R^4$ resp. in $\cN =4$ SYM, these unphysical modes
are eaten up by the massive modes of the gauge bosons via the Higgs effect.
Nevertheless we will add the gauge-fixing term $f^2$ to the potential, because it simplifies the analysis of 
the vector fluctuations. 
This means that the pure gauge zero modes of the 
full vector Laplacian \eq{vector-laplace} become massive, while the physical spectrum is unchanged. 
This should be kept in mind in the 
following analysis.

Hence we consider the following ``gauge-fixed'' potential 
up to quadratic order in $\cA_a$ 
\begin{align}
 V[\cA] & = \tr\Big(\cA_a  \Big(\frac 12\Box_X\d^a_b +  [([X_a, X_b] - 2i g_{abc} X_c),. \, ] \Big) \cA_b \Big) \nn\\
  &= \tr\Big(\cA_a\Big(\frac 12\Box_X \d^a_b  + i c_{abi} [H_{i},.] -  i g_{abc}[X_c ,.]\Big) \cA_b \Big)  \nn\\
  &= V_1[\cA] - V_2[\cA] \ .
 \label{quad-action-fluct-gf}
\end{align}
The last form is shown in appendix \ref{sec:app-pos}, where
\begin{align}
 V_1[\cA] &= \Tr \cA\, \frac 14\Big(\sum_{a=1}^8 [T_a+\l_a,[T_a+\l_a,.]]\Big)\cA       \nn\\
  V_2[\cA] &= \Tr \cA \Big( \frac 14([H_3,[H_3,.]] + [H_8,[H_8,.]])+ ([\l_3, [H_3,.]] + [\l_8,[H_8,.]]) + 3 \Big) \cA 
\end{align}
using the notation
\begin{align}
 \cA = \sum_{a\in  \cI} \l^a \otimes\cA_a  \quad \in \ (8)\otimes End(\cH) .
 \label{cA-vector}
\end{align}
It is not obvious a priori whether or not this potential is stable at the origin, and 
we do not have an analytic expression for the full spectrum. However, 
we  prove in appendix \ref{sec:app-pos} that $V[\cA]$ has no unstable directions.
This is non-trivial, and there are a number of non-trivial zero modes as discussed below.
Our proof is based on a group-theoretical analysis 
in terms of representations of $\msu(3)$, and applies also to arbitrary stacks of 
squashed fuzzy $\cC[\mu_i]$ solutions.  
The results have been verified numerically for the first few  representations.
Thus we have found a large class of non-trivial vacua of 
the model under consideration, with interesting properties discussed in some detail below.

\subsubsection{Vector zero modes}
\label{sec:zero-modes-vector}

There are a number of zero modes or flat directions of the above quadratic action,
which fall into two classes called regular and exceptional zero modes. 
Here we merely state the results, delegating the technical analysis to appendix \ref{sec:app-zeromodes}.

We can decompose the fluctuations into irreducible representations,
\begin{align}
 \cA_a \in End(\cH) = \oplus n_\L \cH_\L .
\end{align}
Using the notation \eq{cA-vector}, 
 regular zero modes are given by 
\begin{align}
 \cA_{\L'} = \l_\rho \otimes v_\L \qquad \in (8) \otimes \cH_\L
 \label{zero-modes-vector-form}
\end{align}
for each summand $\cH_\L \in End(\cH)$, where $\rho = \a_1+\a_2 = \a_3$ is the Weyl vector of $\msu(3)$, 
so that $\l_\rho=\frac 12(\l_1+i\l_2)$ in the standard Gell-Mann basis. 
Further regular zero modes are obtained by acting with the Weyl group on it, 
\begin{align}
 \cA_{w_i\L'} := w_i\cdot \cA_{\L'} = w_i\cdot \l_\rho \otimes w_i\cdot v_\L
\end{align}
for $w_i\in\cW$ acting on $(8) \otimes \cH_\L$. 
Here we recall that the Weyl group -- or more precisely a certain covering in the
braid group -- can be viewed as a subgroup of $SU(3)$, and therefore acts on any 
integrable representation of $\msu(3)$.
It is not hard to verify that these are indeed zero modes, using the last form of \eq{quad-action-fluct-gf}.
There are precisely 6 such regular zero modes for each highest weight vector occurring in the 
decomposition of $End(\cH)$, because $\L' = \L+\rho$
is always in the interior of the fundamental Weyl chamber.

To understand the significance of the regular zero modes, we consider the particular case of 
$\L$ being the maximal weight in $End(\cH_\mu)$.
The corresponding highest weight state $v_\L \in End(\cH_\mu)$ can be written as
\begin{align}
v_{\L} = |\mu\rangle\langle \Omega \mu|
\end{align}
where $\Omega\mu$ is in Weyl chamber opposite to $\mu$. 
This follows from
 \begin{align}
  X_i^+|\mu\rangle &= 0 = \langle \Omega \mu| X_i^+  
 \end{align}
for each $i=1,2,3$. 
Thus $v_{\L}$ maps the lowest weight vector $|\Omega \mu\rangle$  into the highest weight vector $|\mu\rangle$ 
(recall that $-\Omega\mu$ is the highest weight of the conjugate module $\bar\mu$),
and similarly for their Weyl reflections $\cW v_\L$. 
Therefore these particular regular zero modes can be written as
\begin{align}
 \cA_{\L'}  =\l_\rho \otimes |\mu\rangle\langle\Omega\mu| 
\end{align}
and its images under $\cW$.
We will see below that for the squashed $\cC_N^6[\mu]$ branes,
each of the 6 coherent states $\cW|\mu\rangle$ is localized on one of the 3+3 sheets 
at the origin. 
Hence the 6 regular zero modes on  squashed $\cC_N^6[\mu]$ can be interpreted as oriented links (or strings) 
connecting the 3+3 coinciding sheets
with opposite orientation at the origin. 
For the squashed $\cC_N^4[\mu]$  branes, each of the 3 coherent states $\cW|\mu\rangle$ 
is localized on one of  3 mutually intersecting sheets at the origin, and 
the 6 regular zero modes arise from links between different sheets.

Explicitly, these 6 zero modes can be written  as 
\begin{align}
 \cA &= \sum a_{i\pm}\l_i^\pm(X_i^\pm)^{n_{\rm max}}
  \label{linear-deformations}
\end{align}
with maximal $n_{\rm max}$ 
(this is to be contrasted with the background, which has the form $\sum \l_i^\pm X_i^\mp$).
Further regular zero modes corresponding to the other highest weight vectors in $End(\cH_\mu)$ are
given by
\begin{align}
\cA &= \sum a_{i\pm}\l_i^\pm(X_i^\pm)^m, \qquad m < n_{\rm max}
\end{align}
since $(X_i^\pm)^m$ are the extremal weight vectors in $\cH_{(m,m)}\subset End(\cH_\mu)$.
They also link the 3+3 coinciding sheets of $\cC_\cN[\mu]$ 
with opposite orientation, slightly off the origin.  
These exhaust the regular zero modes on $\C P^2_N$ due to \eq{harmonic-decomp-CP2},
but in general there are other regular zero modes arising from highest weights $\L \neq (m,m)$.

Now consider the exceptional zero modes. It turns out (see appendix \ref{sec:app-zeromodes}) that they  
arise for  $\L' = (m,0)$ and $\L' = (0,m)$ in \eq{zero-modes-vector-form}, as well as their
images under the Weyl group. 
There are 3 such exceptional modes for each highest weight $\L = (m-2,1)$ 
and 3  for each $\L = (1,m-2)$
in the decomposition of $End(\cH)$.
Although our analysis in appendix \ref{sec:app-zeromodes} contains a plausible but unproven assumption, we are 
confident that this gives the complete list of zero modes.
Numerically, we find  indeed 6 zero modes for $\L = (0,3), (3,0)$ and $(2,2)$, as expected from the above analysis. 
For $\L = (1,1)$ we find 12 zero modes: 6 regular ones with weights located at the extremal weights 
$\L' = (2,2)$ and its Weyl images, and 6 at the extremal weights $(3,0)$ and $(0,3)$ and their Weyl images.
The latter are identified with the exceptional zero modes.

The geometrical meaning of the exceptional zero modes  is less clear. 
For $\C P^2_N$ with $\mu=(N,0)$ resp. $\mu=(0,N)$, 6 such exceptional zero modes are expected to 
arise due to $\L=(1,1)$ in \eq{harmonic-decomp-CP2}.
For $\mu=(N,1)$ and $\mu=(1,N)$ with $N\geq 2$, 24 such exceptional zero modes are expected to 
arise due to $\L \in\{2\times(1,1), (1,4), (4,1)\}$ in \eq{harmonic-decomp-CP2-S2}.

\subsubsection{Scalar Laplacian}
\label{sec:scalar-Laplace}

The 4-dimensional gauge bosons $A_\mu(x)$ and their masses\footnote{The gauge bosons on $\R^4$ acquire a mass due to 
the Higgs effect, in the non-trivial background of  the scalar fields $X^a$.}
are determined by the scalar Laplace operator $\Box_X$ on 
squashed $\cC_N[\mu]$. Its spectrum is easy to obtain: denoting with $Y^{\L}_{m}$
the states in $\cH_\L \subset End(\cH)$ with weight $m$, we have 
\begin{align}
 \Box_X Y^{\L}_{m} &= 2(\langle\L,\L+2\rho\rangle - \langle m,m\rangle) Y^{\L}_{m} \nn\\
  &= 2(\langle\L+\rho,\L+\rho\rangle - \langle \rho,\rho\rangle - \langle m,m\rangle) Y^{\L}_{m} \nn\\
  &= E_{\L m;s} Y^{\L}_{m} \ .
 \end{align}
 The eigenvalue is the same for any Weyl reflection $\cW m$, so that we can assume that $m$ is in the fundamental Weyl 
 chamber. Then\footnote{The first estimate follows from \eq{weight-inequality} as discussed in appendix
 \ref{sec:app-pos}.}
  \begin{align}
  E_{\L m;s} &= 2(\langle \L +\rho,\L +\rho\rangle - \langle \rho,\rho\rangle - \langle m,m\rangle) \nn\\
  & \geq 4 \langle m,\rho\rangle \geq 0
\end{align}
since $m$ is in the fundamental Weyl chamber. Therefore zero modes are possible only
for $m=0$. However then the first estimate needs to be an equality, which is
possible only for $m=\L$. Therefore there are no non-trivial zero modes of $\Box_X$, and
the only zero mode is the constant function $Y^{0}_{0} \sim \one$ with $\L = 0$.

\section{More fuzzy geometry}
\label{sec:geometry}

To gain more insights into the  geometry of these squashed branes, it is useful to consider coherent states on 
their ``parent`` coadjoint orbits $\cC_\cN(\mu)$. 
They are defined as orbits of the  highest weight states of $\cH_\mu$, 
\begin{align}
 |p\rangle = g_p \cdot |\mu\rangle \ \in \cH_\mu 
\end{align}
where $g_p \mu = p \ \in \cC[\mu]$. 
They are optimally localized in a suitable sense \cite{Perelomov:1986tf},
and allow to extract the semi-classical geometry. 
The  location of such a state on the brane embedded in $\R^8$ is 
\begin{align}
 \langle p|X^a| p\rangle = x^a(p) .
 \label{position}
\end{align}
Note that the action of $g\in SU(3)$ on the lhs is equivalent to the action of $SU(3)$ on 
$\mu\in \mg_0$ in \eq{class-orbit}.
In particular, for the extremal highest weight states $|\mu\rangle \in \cH_\mu$ and their images under $\cW$, 
the corresponding expectation values are
\begin{align}
 \langle \mu | x^{a}| \mu\rangle &= 0, \quad a \in \cI  \nn\\
 \langle \mu | x^{a}| \mu\rangle &= H^a(\mu), \quad a = 3,8 .
\end{align}
Hence the coherent state $|\mu\rangle$ is  localized at $\mu\in\R^8$, 
so that the orbit $x^a(p)$  swept out by the coherent states is indeed $\cC[\mu]$.
After the projection, the extremal states are  located at the origin of $\R^6$,
while the coherent states with weight zero are localized at the boundary 
due to the constraint $C_2 = const$ \eq{C2-C3} on $\cC[\mu]$.

\paragraph{Squashed $\cC^6_N(\mu)$.}

Recall that for $\cC^6_N(\mu)$, the stabilizer group $\cK$ of $\mu$ is 2-dimensional.
Therefore the orbit $g\cdot |\mu \rangle$ for $g$ near the identity $e$
leads to deformations of the position \eq{position} in 6 independent directions, so that
the squashed $\cC^6_N(\mu)$ covers all 6 dimensions near the origin.
The same holds for all images of $\mu$ under the Weyl group $\cW$, which acts freely on $\mu$;
we recall that the Weyl reflections $w_i$ can be lifted to act on
$\mg \cong \R^8$ as elements in $SU(2)_i \subset SU(3)$.
Therefore squashed $\cC^6_N(\mu)$ is a covering  of $\R^6$ near the origin by 6 coincident sheets.
These sheets carry a Poisson or symplectic structure as determined below, which is 
respected by the Weyl group action.

In particular, consider coherent states $|m\rangle_i = g_i \cdot |\mu\rangle$ on the edges
of such a representation $\cH_\mu$,
which are obtained by acting with some $g_i \in SU(2)_i$ on $|\mu\rangle$ (resp.
its images under $\cW$). For these states,
four of the six coordinates $ \langle m | x^{a}| m\rangle_i, \ a\in\cI$   vanish,
and the remaining two combined with a Cartan generator 
$ \langle m |H_{\a_i} | m\rangle_i$ \eq{isomorphism} form a sphere.
Hence these edge states $|m\rangle_i $ correspond to fuzzy spheres through $\mu$,
reflecting the $S^2$ foliation of $\cC^6[\mu]$ near  $\mu$.
In particular, consider the case $\mu=(N,1)$. Then the short edge ending at $\mu$
defines a minimal fuzzy sphere $S^2_{N=2}$, hence $\cC^6_N(\mu)$ should locally look like
$S^2_{N=2} \times \C P^2_N$. This correspondence is borne out at 
the level of representations\footnote{The relation \eq{V-N-1-structure} is shown  in \cite{Grosse:2004wm} 
in a different context, leading to an interpretation as $SU(2)$ instanton bundles over $\C P^2_N$.}, since
\begin{align}
 \cH_{(N,1)} \approx \cH_{(N,0)} \oplus  \cH_{(N+1,0)} \approx  \cH_{(N,0)} \otimes \C^2 ,
 \label{V-N-1-structure}
\end{align}
 Here $\approx$ indicates that the approximation is good 
 for states with weight near $\mu$, where it becomes exact in a suitable sense as $N\to\infty$.
 Thus
\begin{align}
  End(\cH_{(N,1)}) \approx End(\cH_{(N,0)}) \otimes End(\C^2) .
\end{align}
Therefore the ''local fuzzy geometry`` of $\cC^6_N(\mu)$ near $|\mu\rangle$ is
indeed that of $\C P^2_N \times S^2_{N=2}$.
This is just the local geometry of the brane configuration considered in \cite{Steinacker:2014fja} 
in an approach to the standard model.
After the projection, the  $S^2_{N=2}$ becomes a squashed fuzzy disk 
as discussed in section \ref{sec:squashed-sphere}, embedded along a
2-dimensional hyperplane $\R^2$. 
We will also see that the fermionic zero modes arise at the center of these disks.

The boundary of squashed $\cC^6_N(\mu)\subset\R^6$ is
achieved by coherent states with weight 0. These are analogous to the edge states of the
 fuzzy disks.

\paragraph{Squashed $\C P^2_N$.}

\begin{figure}
\begin{center}
 \includegraphics[width=0.45\textwidth]{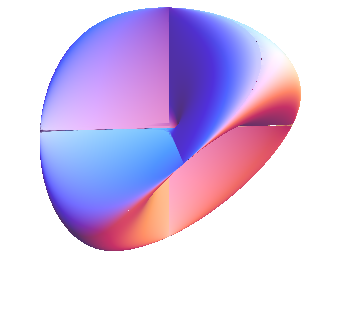}
 \end{center}
 \caption{3-dimensional section of squashed $\C P^2$ through the plane $x_2=x_5=x_7=0$.}
 \label{fig:squashedCP2}
\end{figure}

Consider $\cC^4[\mu]$ for $\mu$ corresponding to $H_8$, which has stabilizer $\cK = SU(2)\times U(1)$. 
The tangent space at $\mu\in \R^8$ is the $4567$ plane, and
the embedding of $\C P^2 \ \hookrightarrow\ \R^8$ near $\mu$
is given explicitly by  \cite{Grosse:2004wm}
\begin{align}
\frac 1{\sqrt{3}}\; (x_1+i x_2) &= \frac{-1}{2x_8+1}(x_4+ix_5)(x_6-ix_7), 
       \nn\\ 
\frac 2{\sqrt{3}}\; x_3 &= \frac{1}{2x_8+1}(x_6^2+x_7^2-x_4^2-x_5^2), 
       \label{constraint-explicit}\\
(1-x_8)(1+2x_8) &= \frac 32 (x_4^2+x_5^2+x_6^2+x_7^2) \nn
\end{align}
taking the branch $x^8 > 0$. 
These equations arise from the $c=1,2,3,8$ components of \eq{CP2-constraints}, the other
constraints are redundant near $\mu$. 
This is also the submanifold given by the coherent states on $\C P^2_N \cong \cC_N[\mu]$ near $\mu$.
Thus after projection to $\R^6$, this sheet of squashed $\C P^2_N$ near $\mu$ 
is embedded along the hyperplane $\R_{4567}$ through the origin. 
The action of the Weyl group leads to 3 such 4-dimensional sheets intersecting at the origin,
along the hyperplanes $\R_{1267},\R_{1245}, $ and $\R_{4567}$.
Therefore squashed $\C P^2_N$ is embedded in $\R^6$ with a triple self-intersection at the origin.
A 3-dimensional section of squashed $\C P^2$ through the plane $x_2=x_5=x_7=0$ is visualized in figure 
\ref{fig:squashedCP2}.
In this picture, the three two-dimensional planes intersecting at the origin are images of the three 
four-dimensional planes. 
One clearly sees how the planes are connected, i.e., there are smooth paths connecting them.
We will see that pairs of fermionic zero modes with both chiralities
arise at the origin, connecting these sheets.

\subsection{Symmetries}

Consider first the global symmetries.
In an 8-matrix model, a potential $\Tr (c_{abc}X^a X^b X^c)$ would admit a $SU(3)$ global symmetry,
transforming the $X^a$ in the $(8)$. On the background of a fuzzy coadjoint orbit
$\cC_N[\mu]$, this symmetry would be spontaneously broken, but 
equivalent to gauge transformations.
After the reduction to the 6-dimensional model with potential $\Tr (g_{abc}X^a X^b X^c)$, 
the global symmetry is reduced to the stabilizer of 
the Cartan generators $H_i$, i.e. the Cartan subalgebra $U(1) \times U(1)$ acting as  
\begin{align}
  X_j^\pm \to X_j^\pm e^{\pm i\varphi_{ij}} 
\end{align}
with suitable phases $\varphi_{ij}$.
In the background of a squashed $\cC_N[\mu]$, this symmetry is again spontaneously broken but
equivalent to gauge transformations,
\begin{align}
  X_j^\pm \to U_i^{-1} X_j^\pm U_i, \qquad
  U_i = e^{i \varphi_{i} H_i} \ .
\end{align}
Therefore this global symmetry 
does not lead to physical zero modes corresponding to Goldstone bosons:
they are eaten up\footnote{In the matrix model without 4-dimensional $\R^4_\theta$ background,
these pure gauge modes are also unphysical, and  absorbed by adding a gauge-fixing term $f^2$ 
as discussed in section \ref{sec:fluctuations}.}
by the massive gauge bosons from the 4-dimensional point of view,
as usual in the Higgs mechanism. Note that 
the $SU(N)$ gauge symmetry is spontaneously broken 
completely in the background of a single brane $\cC_N[\mu]$, with or without projection.

The vector zero modes $\cA = \l_\rho\otimes v_\L$ etc. 
identified in the previous section have non-trivial charge $\L'$ under $U(1)\times U(1)$.
Therefore this remaining global symmetry may be broken if some of these 
bosonic zero modes acquire a non-trivial VEV, and it would no longer be equivalent to 
a gauge transformation.
 
 We also note that the 6 regular $\L=0$ vector zero modes discussed in section \ref{sec:zero-modes-vector}
 can be viewed as Goldstone bosons associated with the translations in the internal 6-dimensional space.
 
 Apart from these continuous symmetries, the background admits the Weyl group $\cW$ of $SU(3)$
 as discrete symmetry, which -- in the absence of fermions -- is again equivalent to gauge transformations.

\subsection{Poisson structure and effective geometry}
 
 So far we discussed the geometry seen by the coherent states, in terms of the target space 
 (or closed string) metric $g_{\a\b}=\del_\a x^a \del_\b x_a$ on $\R^6$.
 However, the effective geometry on noncommutative branes is different, governed by the 
 metric \cite{Steinacker:2010rh}
 \begin{align}
  G^{\a\b} = e^{-\s}\theta^{\a\a'}\theta^{\b\b'}g_{\a'\b'}
  \label{G-eff}
 \end{align}
 where $e^{-\s}$ generalizes the symplectic density $\rho$ in \eq{field-identification}.
 This is the spectral geometry seen by the matrix Laplacian in 
 the semi-classical limit, $\Box_X \sim \Box_G$.
 Since the coadjoint orbits $\cC[\mu]$ are K\"ahler manifolds, their effective fuzzy geometry is the same 
 as the embedding geometry, $G_{\a\b} = g_{\a\b}$ \cite{Steinacker:2010rh}. 
 However after the projection this is no longer the case,
 and indeed the spectrum of $\Box_X$ is clearly not the spectrum of the  
 squashed embedding geometry.
 This can already be seen in the example of the squashed fuzzy sphere, 
 where the embedding metric is flat but it is not the metric underlying $\Box_X$. 
 A  detailed study of this effective geometry is left for future work.
 
We can easily compute the Poisson structure on each sheet of $\cC_N[\mu]$ at the origin,
corresponding to an extremal weight $\cW \mu$. 
It is the push-forward of the $SU(3)$-invariant Poisson structure 
$\{x^a, x^b\} = c^{abc} x^c$
on $\cC[\mu]$ by the projection $\Pi$.
Hence at the origin, the only non-vanishing contributions are
\begin{align}
  \{x^a, x^b\} = c^{abc} t^c = \varepsilon^{abi} H^i[\mu]
\end{align}
on the sheet specified by $\mu$.
Using the explicit structure constants \eq{f-explicit}, this is 
\begin{align}
 \{x^1, x^2\} &= 2 H_3[\mu] = 2(\a_3,\mu)  \nn\\
\{x^4, x^5\} &=  H_3[\mu] + \sqrt{3} H_8[\mu] = 2(\a_1,\mu) \nn\\
\{x^6, x^7\} &= - H_3[\mu] + \sqrt{3} H_8[\mu]  = -2(\a_2,\mu) \ .
\label{poissonbrackets-explicit}
\end{align}
The Poisson structure on the other sheets  corresponding to $\cW\mu$ is different, 
related by the Weyl group. 
This means that the $\Z_3 \subset \cW$  rotation symmetry relating the sheets
with the same orientation
is a discrete family-type symmetry rather than a remnant of a $SU(3)$
gauge symmetry, which would arise on a stack of 3  branes
with the same Poisson structure. Thus the present backgrounds naturally lead to 3 generations.
On $\cC^4[\mu]$, one of the brackets in \eq{poissonbrackets-explicit} of course vanishes.

We can now write down 
the (inverse) orientation form on each sheet of $\cC^6[\mu]$ as given by the Pfaffian of the Poisson structure,
\begin{align}
 \chi &\sim \frac 18  \varepsilon_{abcdef} \{x^a, x^b\} ... \{x^e,x^f\} \
  = {\rm Pf} \theta^{ab} .
\end{align}
This is  the push-forward by $\Pi$
of the inverse (constant) symplectic volume form on $\cC[\mu]$, which is non-trivial, 
and vanishes on the boundary.
At the origin, it is given by
\begin{align}
 \chi &= - (-1)^{|w|}(\a_1,\mu) (\a_2,\mu) (\a_3,\mu)  
\end{align}
where $|w|$ is the signature of the Weyl group element
relating the extremal state corresponding to the sheet as $w|\mu\rangle$.
Hence the orientation of the 3+3 sheets at the 
origin is  determined by the sign of this Weyl group element.
Since the Poisson structure arises on a noncommutative brane carrying a noncommutative $U(1)$ gauge field, 
 the charge of fields linking these branes is determined by this orientation form. 
We will see that the chirality of the fermionic zero modes connecting the sheets
is determined by this orientation, hence by their charge.

\section{Chiral fermions}

The spectrum of 4-dimensional fermions and their masses is governed by the Dirac operator  on 
squashed $\cC_\cN[\mu]$. It can be written as 
\begin{align}
\slashed{D}_{(6)} \Psi = \sum_{a\in\cI} \Delta_a [X_a,\Psi] 
 &= 2\sum_{i=1}^3 \Big(\g_i [X_i^+,.] + \g_i^\dagger [X_1^-,.]\Big) \nn\\
 &= 2\sum_{i=1}^3 \Big(\g_i ad_{T_i^+} + \g_i^\dagger ad_{T_i^-}\Big)
 \label{Dirac-ladder}
\end{align}
where $\Delta_a$ are the 6-dimensional Clifford generators, and 
the spinorial ladder operators
\begin{align}
2\g_1 &= \Delta_4 - i \Delta_5, \qquad 2\g_1^\dagger =  \Delta_4 + i \Delta_5,   \nn\\
2\g_2 &= \Delta_6 + i \Delta_7, \qquad 2\g_2^\dagger =  \Delta_6 - i \Delta_7,   \nn\\
2\g_3 &= \Delta_1 - i \Delta_2, \qquad 2\g_3^\dagger =  \Delta_1 + i \Delta_2,   
\end{align}
 satisfy
\begin{align}
\qquad  \{\g_i,\g_j^\dagger\} = \d_{ij}. 
\end{align}
Our conventions and more details on 
the internal Clifford algebra are given in Appendix \ref{sec:clifford}.
In particular, the partial chirality operator on $\R^2_i$ is  given by 
\bea 
\chi_i &=&  -2(\g_i^\dagger \g_i-\frac 12), 
\label{sub-chirality}
\eea
acting on the spin-$\frac 12$ irreducible representation. 
Using the form \eq{Dirac-ladder}, we can immediately find the zero modes: 
let $v_\L$ be the highest weight vector of $\cH_\L \subset End(\cH_\mu)$. Then 
\begin{align}
\slashed{D}_{(6)} \Psi_{\L} &= 0  
\end{align}
for
\begin{align}
  \Psi_{\L} = |\uparrow\uparrow\uparrow\rangle \otimes v_\L 
  =  |s_1,s_2,s_3\rangle \cdot v_\L \ .
\end{align}
This follows immediately from the decomposition \eq{Dirac-ladder} of the Dirac operator,
noting that
\begin{align}
  \g_i^\dagger|\uparrow\uparrow\uparrow\rangle &= 0 .
\end{align}
Analogous zero modes $\Psi_{w\L} = w \cdot\Psi_\L$ are obtained for any $w\in\cW$:
\begin{align}
  \Psi_{w\L} &= |s_1,s_2,s_3\rangle w\cdot v_\L   \nn\\
    |s_1,s_2,s_3\rangle &= \omega_1 ...\omega_k |\uparrow\uparrow\uparrow\rangle
\end{align}
where $w = w_1 ... w_k$ is a minimal decomposition of $w\in\cW$ 
into elementary reflections along $\a_i$, and 
$\omega_i$ implements the Weyl reflection $w_i$ on the internal spinor space $(\C^2)^{\otimes 3}$
associated to the $\Delta_{1,...,6}$:
\begin{align}
 \omega_i \Sigma_{(i)}  \omega_i^{-1}  &= -\Sigma_{(i)} , \qquad i = 1,2,3 \nn\\
 \omega_i \Sigma_{(j)}  \omega_i^{-1}  &= \Sigma_{(j)} , \qquad j\neq i \nn\\
 \Sigma_{(i)} &= \frac 12[\g_i,\g_i^\dagger] 
  = \frac 12\chi_i \ .
\end{align}
These states have well-defined chirality
\begin{align}
 \Gamma^{(6)}  \Psi_{w\L} &= (-1)^{|w|}  \Psi_{w\L} \ .
\end{align}
There are no other zero modes, since the $\g_i, \g_i^\dagger$ in \eq{Dirac-ladder} are independent.
Recalling the results of section \ref{sec:zero-modes-vector}, we see that 
for each of these zero modes of $\slashed{D}$ there is a corresponding 
regular\footnote{In contrast, the exceptional vector zero modes have no fermionic correspondence, 
 underscoring the fact that SUSY is broken.} zero mode of the 
vector Laplacian, both being in 
one-to-one correspondence to the extremal weights of the irreps in $End(\cH_\mu)$.
They are localized at or near the origin, linking
different sheets with opposite orientation
 on squashed $\cC_N^6[\mu]$, and different 4-dimensional sheets on squashed $\cC_N^4[\mu]$.
For example, the highest weight state $v_\L$ in $End(\cH_\mu)$ can be written as
\begin{align}
v_{\L} = |\mu\rangle\langle \Omega \mu|
\end{align}
where $\Omega\mu$ is in Weyl chamber opposite to $\mu$.

Note that the  chirality of the zero mode $\Psi_{w\L}$
is determined by the sign of the Weyl group 
element $w$. We have seen that this also determines the orientation of the 
corresponding sheets of squashed $\cC_N^6[\mu]$. Therefore the 
chirality of the fermionic zero modes is determined 
by their charges under the noncommutative gauge fields on 
the sheets of squashed $\cC_N^6[\mu]$ linked by $\Psi_{w\L}$. 
Analogous statements hold for  $\cC_N^4[\mu]$.
This can again be illustrated by the gauge field modes $A_\mu^{(3,8)}(x) T_{3,8}$ 
corresponding to the Cartan generators of $\msu(3)$, which are among the lowest of the 
massive Kaluza-Klein gauge field modes according to section \ref{sec:scalar-Laplace}.
They couple to the above zero modes $\Psi_{w\L}$ according to their charge,
which in turn characterizes the different chiralities as shown above,
see figure \ref{fig:8-rep-su3}.
In other words, different chiralities have different gauge couplings.
Of course the total index of the zero modes vanishes and the model 
is guaranteed to be anomaly free
\footnote{Since the gauge group  $U(N)$  is completely broken by the configuration, the chiral
fermion modes of the infrared theory are not coupled to a massless gauge field.
There is no cubic anomaly cancellation that needs to be checked.}, 
presumably leading to some left-right symmetric model in a broken phase. 
Nevertheless,  this is a signature of a  chiral model, 
such as the standard model in the broken phase.
In particular, we note that the 
subgroup $\Z_3$ of rotations in $\cW$ 
preserves the chirality.
This leads to a structure of 3 generations related by $\Z_3$.

Now consider the charge conjugation of these modes. 
First, we have
\begin{align}
v_{\L}^\dagger =  |\Omega\mu\rangle\langle \mu| = \pm\,  v_{\Omega\L} \ .
\end{align}
This is again a zero mode of the above structure, linking the same pair of sheets 
in the opposite sense\footnote{This is the analogous to the fact that the upper-diagonal
zero modes for intersecting 
branes are the charge conjugates of the lower-diagonal ones \cite{Chatzistavrakidis:2011gs}.}.
By an appropriate choice of $\Gamma$ matrices, \cf Appendix~\ref{sec:clifford}, it is straightforward to see that the six-dimensional charge conjugation matrix acts as $C^{(6)} |\uparrow\uparrow\uparrow\rangle = - |\downarrow\downarrow\downarrow\rangle$. Hence, we have pairs $\chi_\pm$ of zero modes of the internal Dirac operator, fulfilling
\begin{align*}
 \Gamma^{(6)} \chi_\pm & = \pm \chi_\pm, &
 C^{(6)} \chi_\pm^* & = \chi_\mp.
\end{align*}
As discussed in Appendix~\ref{sec:clifford}, this entails that we may construct Majorana-Weyl solutions to the full Dirac operator of the form
\[
 \Psi = \psi_+ \otimes \chi_+ + \psi_- \otimes \chi_-,
\]
where the four-dimensional spinors $\psi_\pm$ fulfill
\begin{align}
 \Di_{(4)} \psi_\pm & = 0, &
 \gamma_5 \psi_\pm & = \pm \psi_\pm, &
 \psi_\pm^C & = \psi_\mp.
 \label{MW-explicit}
\end{align}
These fermionic degrees of freedom can be 
viewed either in terms of a 4-dimensional Weyl or Majorana spinor.

In addition to these zero modes, there are  8 trivial fermionic 
zero modes which arise from constant spinors  $\sim \one_N$ i.e. $\L=0$.
The Weyl group acts trivially on these, so there is no family structure, and these fermions are 
non-chiral.
Combined with the  6 trivial vector zero modes (corresponding to 6
space-time scalars) and the trivial scalar zero mode (corresponding to the
2 physical $U(1)$ space-time gauge boson modes) these form a $\cN=4$
SUSY multiplet. This captures the trace-$U(1)$ sector which is trivial in $\cN=4$ SYM,
but acquires an interesting geometrical role related to gravity in the matrix model \cite{Steinacker:2010rh}.

These considerations apply to any irrep in the decomposition 
\begin{align}
 \Psi \in  End(\cH_\mu) \otimes \C^8 = \Big(\oplus_\L \cH_\L  \Big) \otimes \C^8
\end{align}
Therefore there are 3 chiral zero modes (or equivalently their charge conjugates modes)
for each non-trivial component in $End(\cH_\mu)$,
as well as the 8 constant spinors arising from $\L=0$.

\section{Minimal squashed orbits}

To illustrate the above results, we briefly discuss the minimal squashed fuzzy spaces,
with Hilbert space $\cH$ of minimal dimension.

\paragraph{Minimal squashed $\C P^2_{N=3}$.}

This is defined for\footnote{Analogous considerations apply for $\mu=\L_2$.} 
$\mu = \L_1 = (1,0)$. Denote the weights of $\cH_\mu$ with $\mu_1, \mu_2, \mu_3$.
Then 
\begin{align}
 End(\cH_{\mu}) \cong (\L_1) \otimes (\bar\L_1) = (1,1) \oplus (0,0) ,
\end{align}
which is the space of functions on minimal squashed fuzzy $\C P^2_{N=3}$. 
Near the origin, there are 
3 four-dimensional sheets labeled by the extremal states $|\mu_i\rangle$.
Any two pairs of these states form a minimal 
squashed fuzzy disks arising from the $SU(2)_i$ associated with the roots $\a_i$,
embedded in $\R^2_{12}$, $\R^2_{45}$, and $\R^2_{67}$ respectively.
The background admits  18 flat directions, associated with the 6 regular and 6 exceptional vector zero modes from
$\cH_{(1,1)}$ and the 6 trivial flat directions associated with translations.
There are $6=3+3$ fermionic zero modes $\psi\sim |\mu_i\rangle\langle \mu_j|$ for $i \neq j$, 
which arise from the intersections of these 4-dimensional sheets, along e.g. 
$\R^4_{1245} \cap \R^4_{1267} \cong \R^2_{12}$ etc. The intersections consist of two opposite sheets 
of the minimal disk connecting the pair of extremal states $|\mu_i\rangle$.

We note again that the fermionic zero modes $\psi\sim |\mu_i\rangle\langle \mu_j|$
and $\psi\sim |\mu_j\rangle\langle \mu_i|$ have opposite chirality, however they are  
charge conjugates of each other and therefore not independent due to \eq{MW-explicit}. As discussed before, 
this means that the 4-dimensional fermions arising from these zero modes comprise 3 chiral fermions 
(or equivalently their charge conjugate modes), 
corresponding to a low-energy theory with a 3-fold family structure.

\paragraph{Minimal squashed $\cC^6[\mu]$.}

\begin{figure}
\begin{center}
 \includegraphics[width=0.45\textwidth]{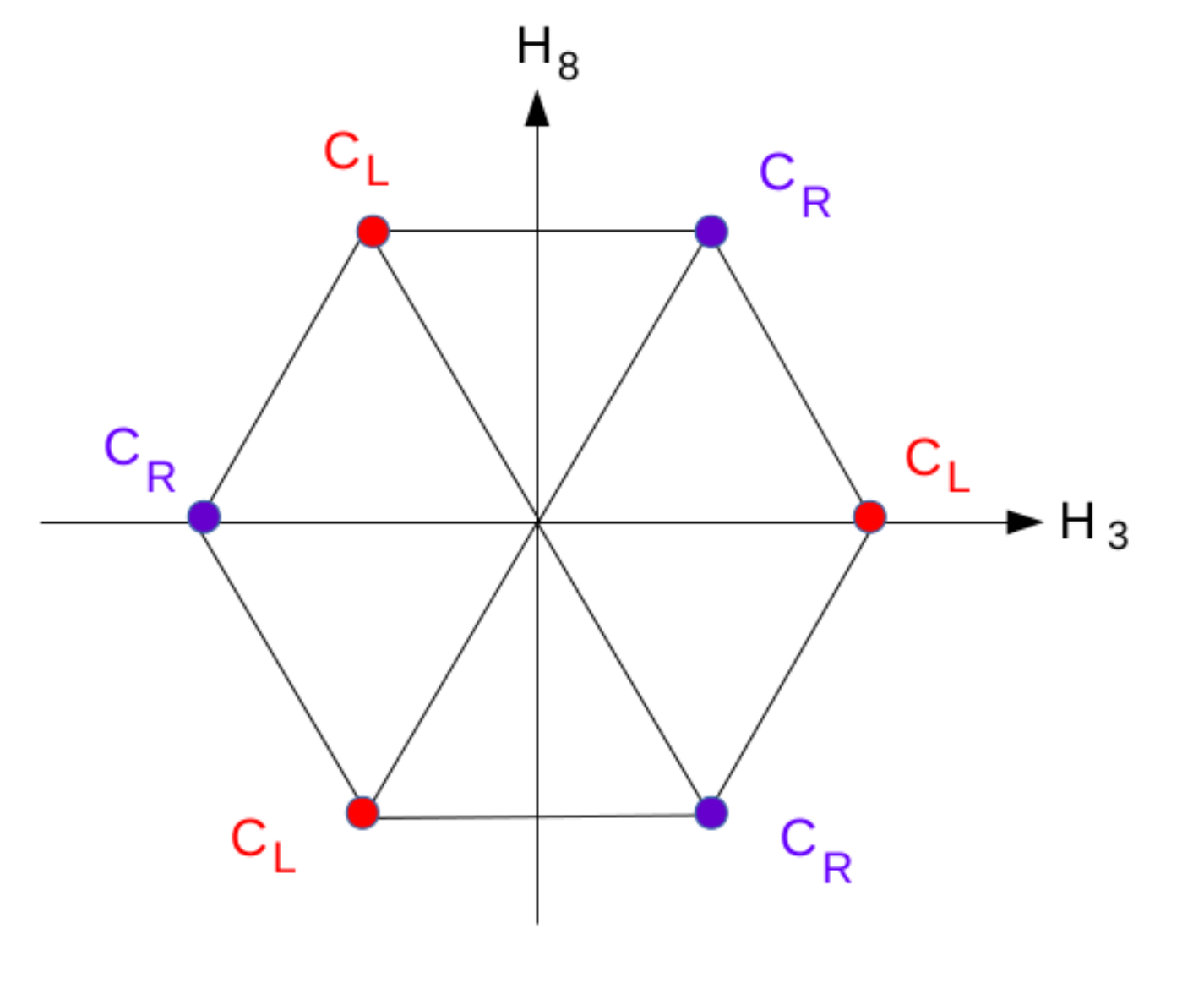}
 \end{center}
 \caption{Decomposition of squashed $\cC[(1,1)]$ into $3+3$ oriented sheets $C_L, C_R$ related by $\Z_3$.}
 \label{fig:8-rep-su3}
\end{figure}

This is defined for $\mu=(1,1)$.
In contrast to  $\cC^4[\mu]$,
the 6 extremal weights of $H_\mu \cong (8)$ now decompose into two sets of 3 weights 
related by the Weyl rotations $\Z_3$. They correspond to  3+3 sheets  $C_L$ resp. $C_R$
with  opposite orientation, as sketched in figure \ref{fig:8-rep-su3}.
The geometry is locally the product of a fuzzy disk with a sheet of squashed $\C P^2_{N=3}$.
This is reminiscent of the geometry considered in \cite{Steinacker:2014fja} in an approach to the standard model,
however now with 3 families.
The space of functions decomposes into
\begin{align}
 End(\cH_{(1,1)}) = (1,1)\otimes (1,1) = (2,2) \oplus (3,0)\oplus (0,3) \oplus 2(1,1) \oplus (0,0) \ .
\end{align}
This leads to 5 sets of fermionic zero modes in 3 generations,
as well as the 8 constant spinors arising from $(0,0)$.
Altogether we expect
$5*6+8 = 38$ fermionic zero modes, consistent with numerical checks.

\section{Stacks of branes}

Now consider solutions of the model corresponding to stacks of 
such squashed $\cD_i = \cC[\mu_i]$ branes corresponding to a reducible representation
of $\msu(3)$, e.g.
\begin{align}
 X^a = \begin{pmatrix}
        X^a_{\mu_1} & \\
        & X^a_{\mu_2}
       \end{pmatrix} .
 \label{block-config}
\end{align}
The off-diagonal blocks decompose into irreducible representation of $\msu(3)$ according to
$Hom(\cH_{\mu_1},\cH_{\mu_2}) \cong \oplus n_\L \cH_\L$. 
In particular, the vector fluctuations are still
governed by the quadratic action \eq{quad-action-fluct-gf}, 
replacing $X_a$ by the reducible generators in \eq{block-config}.
Therefore our positivity result 
applies, and the bosonic fluctuations around this background are non-negative. 
Similarly, the results for the fermionic zero modes and their chirality
carry over immediately for the off-diagonal 
blocks of fermions on such a stack of squashed $\cC[\mu_i]$ branes.

\paragraph{Intersections with point branes.}

In particular, consider a stack consisting of $\cD_1 = \cC[\mu]$ and a point brane $\cD_2 =\cC[0]$.
Then $Hom(\cH_\mu,\C) \cong \cH_\mu$, and
our general results imply that there are off-diagonal fermionic
zero modes arising from the extremal weight states on $\cC[\mu]$, 
connecting $\cD_1$ with $\cD_2$. 
For $\cC^6[\mu]$, these zero modes with opposite chirality
arise on different sheets, so that they may couple effectively to different gauge fields.
For $\mu = (N,1)$ or $\mu = (1,N)$, the extremal
coherent states on $\cC^6[\mu]$ correspond precisely to fermions on 
the (squashed) intersecting branes $S^2_2 \cap \C P^2_N$,
due to the local factorization \eq{V-N-1-structure}.
Therefore the structure of these fermions is completely analogous to those studied on intersecting branes in 
\cite{Steinacker:2014fja} in an approach to the standard model, but now with 3 generations.

\paragraph{Stacks of minimal branes.}

As a second example, 
consider a stack of $n$  minimal fuzzy squashed branes $\cD_{i} \cong \C P^2_{N=3}$. 
This leads to a $U(n)$ gauge symmetry. The massless 4-dimensional fields on this background consist of 
3+3 chiral fermions in the adjoint
with a $\Z_3$ family symmetry, 8 trivial fermions,  18
massless scalars (due to 6 regular and 6 exceptional flat vector directions associated with $\cH_{(1,1)}$
and the 6 trivial flat directions) and one unbroken $U(n)$ gauge field (due to the trivial 
eigenmodes of $\Box_X$). 
The trivial modes clearly form a $\cN=4$ supermultiplet. 
The regular extremal modes can be combined with the fermionic zero modes into
chiral $\cN=1$ supermultiplets, but supersymmetry is manifestly broken  by the  exceptional zero modes.

More interesting physics could be obtained by
switching on some Higgs arising from the vector zero modes. 
The trivial mode $(0,0)\subset (1,0)\otimes (0,1)$ on a background of two coinciding branes
does not break any symmetry, since it shifts both branes.
However the other flat directions will
break the $U(n)$ gauge symmetry, and in general also the family symmetry.
It is then quite conceivable that the 
off-diagonal modes in $Hom(\cD_i,\cD_j)$ lead to interesting low-energy physics.

\section{Energy}

Let us compute the energy of our configurations, i.e.,
\[
 E = - S = - \frac 14\Tr \left( [X_a, X_b] [X_a, X_b] + i f_{abc} X_a [X_b, X_c] \right).
\]
Using the definition \eqref{eq:f_abc}, we obtain 
\begin{align*}
 E & = \frac 14\Tr \left( c_{abi} c_{abj} T_i T_j + g_{abc} g_{abd} X_c X_d + \tfrac{8}{3} i g_{abc} X_a [X_b, X_c] \right) \\
 & = - \frac 14 c_N \left( \tfrac{10}{3} g_{abc} g_{abc} - 2 c_{abi} c_{abi} \right)  \nn\\
 & = - 8 c_N 
\end{align*}
where we used the notation introduced in \eqref{eq:def_g}. For the expression in brackets 
in the 2nd line, one obtains $\frac{10}{3} \times 6 \times 4 - 2 \times 2 \times 12 = 32$,
using the structure constants \eqref{f-explicit}. 
This shows that our configurations has indeed lower  energy than the trivial solution. 
We note that upon a rescaling $X_a \to \lambda X_a$, also $f_{abc}$ has to scale, $f_{abc} \to \lambda f_{abc}$, 
so that the energy scales as $E \to \lambda^4 E$.

For an irreducible brane $\cC[\mu]$, this gives 
\begin{align}
 c_N[\mu] &= \frac 1{16} \tr \sum_{a=1}^8 T_a T_a 
  = \frac{\dim(\cH_\mu)}{8}(\mu,\mu+2\rho)  \nn\\
  &= \frac{\dim(\cH_\mu)}{12}(m_1^2 +  m_2^2 +  m_1 m_2 + 3m_1+3m_2) 
\end{align}
for $\mu = m_1 \L_1 + m_2 \L_2$. 
Using Weyls dimension formula, 
the representation space of this brane has dimension
\begin{align}
 \dim \cH_\mu = \frac 12 (m_1+1) (m_2+1) (m_1+m_2+2) .
\end{align}
For minimal squashed $\C P^2_{N=3}$, we obtain 
$c_N[(1,0)] =\frac{3}{12}(1 + 3) = 1$.

One may now ask the following question: for a given Hilbert space $\C^N$, 
which partition into irreducible representations $\C^N = \oplus_\mu n_\mu \cH_\mu$
minimizes the above energy $E = -8\sum n_\mu c_N[\mu]$. This is an interesting discrete optimization problem,
which should uniquely select a brane configuration. 
In particular, if $N$ is {\em not} the dimension of an irreducible representation of $\msu(3)$,
then the minimum among the above configurations is necessarily assumed for a reducible
representation, corresponding to a non-trivial stack of various fuzzy branes $\cC[\mu]$. 
A similar problem was discussed in \cite{Aschieri:2006uw} in a simpler setting.
Quantum corrections might of course modify the conclusion, but in any case 
the model should select its preferred ground state for given flux terms as above. 


It is instructive to assume a single brane and to allow for real $m_i \geq 0$. 
Then the extremum of $c_N[\mu]$ for fixed $\dim \cH_\mu$ is attained at $m_1 = m_2 = (\dim \cH_\mu)^{\frac{1}{3}} - 1$, 
which is a minimum. Hence, the maximum, and thus the minimal energy, is attained for the degenerate 
configuration where one of the $m_i$ vanishes. 
The (almost) degenerate configurations with one large and one small $m_i$ thus seem to be dynamically preferred.

\paragraph{Fuzzy sphere solution.}

Besides these squashed $\cC[\mu]$ solutions, there are also the 
well-known fuzzy sphere solutions discussed in section \ref{sec:fuzzy-sphere}. 
Recalling the explicit form for $f_{abc}$ \eqref{eq:f_abc} and using \eqref{S2-solution},
these are given e.g. by
\begin{align}
 X_4 &= T_1, \qquad 
 X_6 = T_2,  \qquad
 X_1 = T_3 
\end{align}
and $X_2 = X_5 = X_7 = 0$, or more compactly
\begin{align}
 X_1^\pm = \frac 12 T_1, \ \  X_2^\pm = \frac 12 T_2, \ \   X_3^\pm = \frac 12  T_3 \ .
\end{align}
Of course there are analogous solutions by introducing phases to the $X_i^\pm$.
Their energy is 
\begin{align*}
 E & = \frac 14\Tr \left(\sum_i 8 T_i T_i - i f_{abc} X_a [X_b, X_c] \right) \\
 & = \frac 12 3 c_N \left(8 - \frac {32}3 \right)  
 = - 4 c_N 
\end{align*}
where 
\begin{align}
 c_N = \frac 16 \Tr (\sum_i T_i T_i) = \frac 16 N(N^2-1) 
\end{align}
and $N = \dim \cH$. Hence 
the energy of the fuzzy sphere for $N=3$ is
$E=-16$, which is lower than that of $\C P^2_{N=3}$. For larger $N$, the 
difference will be even more pronounced.
However quantum corrections will of course modify this result, and 
the massless modes on $\cC_N[\mu]$ may play a significant role. 
Since there are no negative modes on $\cC_N[\mu]$, it is 
quite conceivable that these backgrounds will become at least local minima. 
This should be studied in detail elsewhere.

\section{Conclusion and outlook}

We discussed a new class of vacuum solutions of $\cN=4$ SYM in the presence of suitable 
cubic terms in the potential. These brane solutions have the geometry of 
4- and 6-dimensional projected or squashed 
coadjoint orbits of $SU(3)$, and naturally lead to
3 generations of fermionic zero modes with distinct chirality properties
determined by their $\msu(3)$ charges.
The bosonic and fermionic zero modes were studied in detail.

This work originated from the observation that besides the 
well-known fuzzy sphere solutions, the $\cN=4$ action with cubic terms admits
many  other non-trivial solutions, 
such as intersecting fuzzy spheres and similar geometries; we hope 
to report on these elsewhere. Such intersecting brane solutions 
are interesting because they may lead to low-energy physics close to the standard model
\cite{Steinacker:2014fja}. However, these solutions typically have some negative modes.
It is therefore very remarkable that 
the new vacua under consideration here have no negative modes but a number of flat directions, which we 
identify explicitly. These vacua are therefore promising starting points for 
further investigations.

There are many  issues which should be addressed in further work. 
One important questions is whether the massless modes are stabilized 
by quantum corrections, and whether some of them acquire a non-trivial VEV. 
This question could be addressed within a weak coupling analysis.
The effect of adding an explicit mass term should also be worked out in this context.

Another question is to clarify the effects of 
turning on nontrivial VEVs along some of these flat directions.
For 
stacks of such brane solutions, this  might lead to interesting low-energy physics,
due to the similarity with the geometry considered in
 \cite{Steinacker:2014fja} in an approach to the standard model.
Therefore a more detailed understanding of the bosonic flat directions and their
possible stabilization is very important to assess the physical significance of the 
new vacua. This is a non-trivial problem both classically as well as at the quantum level.
It would also be important to understand the effects of modifying the cubic potential. 
In the context of string theory, a natural question is whether these backgrounds have a 
dual descriptions in terms of supergravity, which might help to shed light on  the 
strong coupling regime. 
We emphasize again that the present solutions make perfect sense at weak coupling in the gauge theory, and 
their interpretation in terms of higher-dimensional geometry does not rely on any
holographic picture. 
In any case, it should be clear that $\cN=4$ SYM with soft SUSY breaking terms 
provides a remarkably rich basis for further investigations along these lines.

\paragraph{Acknowledgments.}

This work is supported by the Austrian Science  Fund (FWF): P24713.

\appendix

\section{Clifford algebra and reduction to 4 dimensions}
\label{sec:clifford}

For the $\Gamma$ matrices, we use the conventions of \cite{VanProeyen}.
The ten-dimensional Clifford algebra, generated by
$\Gamma_A$, then naturally separates into
a four-dimensional and a six-dimensional one as follows,
\bea
\Gamma_A&=&(\Gamma_{\mu},\Gamma_{3+a}), \nn\\ \Gamma_{\mu}&=&\gamma_
{\mu}\otimes\one_8, \qquad \Gamma_{3+a} = \gamma_5\otimes\Delta_a.
\eea
Here the $\gamma_{\mu}$ define the four-dimensional Clifford algebra, 
the $\Delta_a$ define the six-dimensional Euclidean Clifford algebra. 
The ten-dimensional chirality operator 
\be
\Gamma = \gamma_5 \otimes\Gamma^{(6)}
\ee
separates into four- and six-dimensional chirality operators 
\bea
\gamma_5 &=& -i\gamma_0 ... \gamma_3 = \gamma_5^\dagger, \nn\\
\Gamma^{(6)} &=& -i\Delta_1 ... \Delta_6 = (\Gamma^{(6)})^\dagger .
\label{Gamma-int}
\eea
Let us denote the ten-dimensional charge conjugation matrix as
\be
\cC = C^{(4)} \otimes C^{(6)}, 
\ee
where $C^{(4)}$ is the four-dimensional charge
conjugation matrix and $C^{(6)}$ the one on $\R^6$. In the basis of \cite{VanProeyen}, we have
\begin{align*}
 \Delta_4 & = \sigma_1 \otimes \one \otimes \one &
 \Delta_6 & = \sigma_3 \otimes \sigma_1 \otimes \one & 
 \Delta_8 & = \sigma_3 \otimes \sigma_3 \otimes \sigma_1 \\ 
 \Delta_5 & = \sigma_2 \otimes \one \otimes \one &
 \Delta_7 & = \sigma_3 \otimes \sigma_2 \otimes \one & 
 \Delta_9 & = \sigma_3 \otimes \sigma_3 \otimes \sigma_2 
\end{align*}
and
\[
 C^{(6)} = \sigma_2 \otimes \sigma_1 \otimes \sigma_2.
\]
Furthermore, ${C^{(4)}}^{-1} = C^{(4)}$ and ${C^{(6)}}^{-1} = C^{(6)}$.
The charge conjugation matrix satisfies, as usual, the relation
$\cC\Gamma^M\cC^{-1}=-(\Gamma^M)^T$. 
Then
the Majorana condition in 9+1 dimensions is 
$\Psi = \Psi^C = \Gamma^0 \cC \Psi^*$,
where $\Psi^*$ denotes the spinor with conjugated matrix entries. In particular, the charge conjugation factorizes as
\[
 (\Psi^{(4)} \otimes \Psi^{(6)})^C = \gamma^0 C^{(4)} \Psi_{(4)}^* \otimes C^{(6)} \Psi_{(6)}^*.
\]
The fermionic action can then be written as
\begin{align}
 \Tr \obar\Psi \Gamma_a[X^a,\Psi] &= -\Tr \Psi^T \cC^T \big(\slashed{D}_4 + \gamma_5\slashed{D}_{(6)}\big)\Psi
\end{align}
where
\begin{align}
\slashed D_{(6)} = \sum_{a\in \cI}\Delta^a [X_a,.], 
\qquad 
\{\slashed D_{(6)},\Gamma^{(6)}\} =0 \ .
\label{D-6-chirality}
\end{align}
Here $\cI$ denotes the internal indices, and $\slashed D_{(6)}$
denotes the Dirac operator on the internal space.
Assume that we have a pair $\chi_\pm$ of zero eigenvectors of $\Di_{(6)}$, with chirality $\Gamma^{(6)} \chi_\pm = \pm \chi_\pm$, and related by $C^{(6)} \chi_+^* = \chi_-$. Then
\[
 \Psi = \psi_+ \otimes \chi_+ + \psi_- \otimes \chi_-
\]
is a Majorana-Weyl spinor iff $\gamma_5 \psi_\pm = \pm \psi_\pm$ and $\psi_+^C = \gamma^0 C^{(4)} \psi_+^* = \psi_-$. Furthermore, it is a solution to the Dirac operator iff $\Di_{(4)} \psi_\pm = 0$.


\section{Positivity of the vector fluctuations}
\label{sec:app-pos}

Consider a background $X^a = \pi(T^a)$ for $a \in \cI$, for any finite-dimensional unitary representation $\pi$ 
(not necessarily irreducible) of $\msu(3)$ acting on $\cH$. Let $\cA^a$ be the fluctuations around this background.
We will  show that the potential of these vector fluctuations  \eq{quad-action-fluct-gf}
\begin{align}
 V[\cA] &= \tr\Big(\cA_a^\dagger\Big(\frac 12\Box_X\cA_a + i c_{abi} [H_{i},.]\Big) \cA_b \Big)  
   - \tr\Big(\cA_a^\dagger i g_{abc}[X_c ,\cA_b ] \Big)  \nn\\
  &= V_\Box[\cA] + V_H[\cA]  + V_g[\cA]
\end{align}
is non-negative. For technical reasons we do not require $\cA_a$ to be hermitian.
 We separate this potential into three parts,
 using the notation \eq{cA-vector}
\begin{align}
 \cA = \sum_{a\in  \cI} \l^a \otimes\cA_a  \quad \in \ (8)\otimes End(\cH) ;
 \label{cA-notation}
\end{align}
observe that the Cartan generators are missing, which amounts to a constraint on $\cA$.
First,
\begin{align}
 V_\Box[\cA] &= \frac 12\tr\sum_a\cA_a^\dagger \Box_X\cA_a
    = \frac 14 \Tr\cA^\dagger \Box_X\cA  .
\end{align}
Furthermore
\begin{align}
V_g[\cA]  &= i \sum_{a,b,c \in\cI} g_{abc} \tr(\cA_a^\dagger [X_b, \cA_c]) \nn\\
   &= \frac 12\sum_{a,b,c \in\cI} \Tr(\l_a [\l_b,\l_c]  \cA_a^\dagger [X_b, \cA_c]) \nn\\
   &=  \frac 12\sum_{b,c \in\cI}\Tr(\cA^\dagger [\l_b, [X_b, \cA_c\l_c]]) \nn\\
  &= \frac 12\Tr\cA^\dagger\slashed{D}_{ad(6)} \cA
\end{align}
where we define the ``adjoint'' Dirac operator
\begin{align}
 \slashed{D}_{ad(6)} \cA = \sum_{b\in\cI} [\l^b,[X^b, \cA]]
\end{align}
acting on $\cA\in (8)\otimes \cH_\L$.
Similarly, using \eq{epsilon-Cartan-rel} we can write
\begin{align}
 V_H[\cA] &= \sum_{a,b \in\cI}\tr\Big(\cA_a^\dagger i c_{abi} [H_{i},.] \cA_b \Big)  \nn\\
  &= -\frac 12\sum_{a,b \in\cI}\sum_{i=3,8}\Tr\Big(\l^a\cA_a^\dagger  [\l_{i},\l_b] [H_{i},\cA_b]  \Big) \nn\\
  &= -\frac 12 \sum_{i=3,8} \Tr\Big(\cA^\dagger [\l_{i},[H_{i},\cA]]   \Big) .
\end{align}
Therefore the fluctuations are governed by the potential
\begin{align}
 V[\cA] &= \Tr \cA^\dagger \Big(\frac 14\Box_X  + \frac 12\slashed{D}_{ad(6)} - \frac 12([\l_3, [H_3,.]] + [\l_8,[H_8,.]]) \Big) \cA \nn\\
 &= \Tr \cA^\dagger \Big(\frac 14\Box_T  + \frac 12\slashed{D}_{ad(8)} 
  - \frac 14([H_3,[H_3,.]] + [H_8,[H_8,.]]) - ([\l_3, [H_3,.]] + [\l_8,[H_8,.]]) \Big) \cA \nn\\
  &= \Tr \cA^\dagger \Big(\frac 14\big(\Box_{T+L}  - \sum_{a=1}^8 L_a L_a\big)
  - \frac 14([H_3,[H_3,.]] + [H_8,[H_8,.]]) 
   - ([\l_3, [H_3,.]] + [\l_8,[H_8,.]])  \Big) \cA \nn\\
  &= V_1(\cA) - V_2(\cA) 
\end{align}
denoting $L^a = [\l^a,.]$ acting on $(8)$, where
\begin{align}
  V_1[\cA] &= \Tr \cA^\dagger\, \frac 14\Box_{T+L}\cA \nn\\
  V_2[\cA] &= \Tr \cA^\dagger \Big( \frac 14([H_3,[H_3,.]] + [H_8,[H_8,.]])+ ([\l_3, [H_3,.]] + [\l_8,[H_8,.]]) + 3 \Big) \cA \nn\\
 \Box_{T} &= \sum_{a=1}^8 [T_a,[T_a,.]]   \nn\\
  \Box_{T+L} &= \sum_{a=1}^8 ([T_a,.] + L_a) ([T_a,.] + L_a) .
\end{align}
Here we used
 \begin{align}
 \sum_{a=1}^8 \frac 12 L_a L_a &= (\rho,3\rho) = 6   .
\end{align}
Now decompose 
\begin{align}
 End(\cH) = \oplus\, m_\L \cH_\L 
\end{align}
where $\cH_\L$ denotes a highest weight irrep with highest 
weight $\L$. Since the operators $\cO_1, \cO_2$ defining 
$V_{1,2} = \Tr \cA^\dagger \cO_{1,2}\cA$ involve only Lie algebra generators, 
it is enough to show positivity for $\cA_a \in \cH_\L$ with fixed $\L$, and
we can focus on any irreducible component $\cH_\L$.
Consider the ``composite'' fluctuation
\begin{align}
 \cA \ \in (8) \otimes  \cH_\L  = \oplus_j n_j \cH_{\L+L_j} = \oplus_j n_j \cH_{\L'_j}
\end{align}
where 
\begin{align}
 \L'_j = \L + L_j ,
\end{align}
and the sum is over the weights $L_i$ of\footnote{This follows from the character decomposition
or equivalently the Racah-Speiser algorithm. } $(1,1)$; the multiplicities $n_j$ are one 
or zero for $L_j\neq 0$, and possibly two for $L_j=0$. 
Since both operators $\cO_1, \cO_2$ defining 
$V_{1,2} = \Tr \cA^\dagger \cO_{1,2}\cA$ commute with $L_3 + [X_3,.]$ and $L_8 + [X_8,.]$, they respect
the overall weight $M'$ in $(8)\otimes \cH_\L$, and different $M'$ contributions are orthogonal.
It is therefore sufficient to prove positivity for $\cA \equiv \cA_{M'}$, 
and we fix $M'$ from now on.
Furthermore, the constraint \eq{cA-notation} on $\cA$ means that we can write
\begin{align}
 \cA_{M'} = \sum_i \cA_i = \sum_i  \l_{L_i}\otimes  \cA^i_{M_i}, \qquad L_i\neq 0, \ \ M_i+L_i = M' \ .
 \label{Mprime-expand-tensor}
\end{align}
Moreover $\cO_2$ respects both $L$ and $M$. 
Therefore
\begin{align}
  V_2[\cA]  &= \sum_i\Tr \cA_i^\dagger \Big(\frac 12(M_i,M_i) +2(M_i,L_i) +3 \Big) \cA_i \nn\\
  &= \sum_i\Tr \cA_i^\dagger \Big( \frac 12(M_i+2L_i,M_i+2L_i) - 2(L_i,L_i) +3 \Big) \cA_i \nn\\
  &= \sum_i\Tr \cA_i^\dagger \Big( \frac 12(M_i+2L_i,M_i+2L_i) - 1 \Big) \cA_i  \nn\\
  &= \sum_i\Big( \frac 12(M_i+2L_i,M_i+2L_i) - 1 \Big)\Tr \cA_i^\dagger  \cA_i 
  \label{S-2}
\end{align}
recalling that $(L_i,L_i) = 2$ due to the constraint $L_i\neq 0$.
Now consider $V_1[\cA]$. It is now useful to use the decomposition 
\begin{align}
 \cA &= \sum_{\L'_j} \cA_{\L'_j}^{M'} \ \in \  \oplus_{\L'_j}  \cH_{\L'_j}
 \label{LM-L'-decomp}
\end{align} 
Clearly $M'$ is a weight in each of the $\cH_{\L'}$ occurring in this sum. 
Now for any $\cA_{\L'}^{M'}  \in  \cH_{\L'}$ with $M'$ being a weight of $\cH_{\L'}$, we have
\begin{align}
  V_1[\cA_{\L'}^{M'}] &= \Tr (\cA_{\L'}^{M'})^\dagger\frac 12 (\L',\L'+2\rho) \cA_{\L'}^{M'} \nn\\
   &= \Tr (\cA_{\L'}^{M'})^\dagger \big(\frac 12(\L'+\rho,\L'+\rho) -1\big)\cA_{\L'}^{M'} .
     \label{S-1}
\end{align}
Here we observe that the Weyl vector for $\msu(3)$ is given by 
$\rho = \sum_{i=1,2}  \L_i = \sum_{i=1,2} \a_i = \a_3$ with $(\rho,\rho) = 2$. 
Therefore 
\begin{align}
 V_1[\cA] &= V_1[\sum_{\L'} \cA_{\L'}^{M'}]  
  = \sum_{\L'} \big(\frac 12(\L'+\rho,\L'+\rho) -1\big) \Tr ((\cA_{\L'}^{M'})^\dagger \cA_{\L'}^{M'}) 
\label{S1-estimate}
\end{align}
because different irreps are orthogonal, 
$\Tr ((\cA_{\L'}^{M'})^\dagger \cA_{\L''}^{M'}) = \d_{\L'\L''} \Tr ((\cA_{\L'}^{M'})^\dagger \cA_{\L'}^{M'})$.
Our strategy is to show that 
\begin{align}
 (\L'+\rho,\L'+\rho) -2 &\geq (M'+\rho,M'+\rho) -2 \geq   (M_i+2L_i,M_i+2L_i) - 2             
  \label{estimate}
\end{align}
for each $\L'_i$ in this sum, and each $(M_i,L_i)$ occurring in \eq{S-2}. 
If this is satisfied, then we can continue \eq{S1-estimate} as follows
\begin{align}
 V_1[\cA] &= \sum_{\L'} \big(\frac 12(\L'+\rho,\L'+\rho) -1\big) \Tr ((\cA_{\L'}^{M'})^\dagger \cA_{\L'}^{M'}) \nn\\
  &\geq \sum_{\L'} \big(\frac 12(M'+\rho,M'+\rho) -1 \big) \Tr ((\cA_{\L'}^{M'})^\dagger \cA_{\L'}^{M'}) \nn\\
   &= \big(\frac 12(M'+\rho,M'+\rho) -1 \big) \sum_{\L',\L''} \Tr ((\cA_{\L'}^{M'})^\dagger \cA_{\L''}^{M'})  \nn\\
    &= \big(\frac 12(M'+\rho,M'+\rho) -1 \big)  \Tr (\cA^\dagger \cA) \nn\\
    &\geq \sum_i\Big( \frac 12(M_i+2L_i,M_i+2L_i) - 1 \Big)\Tr \cA_i^\dagger  \cA_i \nn\\
    &=  V_2[\cA]
\label{S1-estimate-2}
\end{align}
using \eq{S-2}, as desired.
To show \eq{estimate}, we first observe that $M'=M_i+L_i$ is a weight in each $\cH_{\L'}$ in \eq{S1-estimate}.
Then we can use the following lemma (see e.g. \cite{jacobson}  Chapter VIII lemma 3):
Let  $\L'$ be a dominant weight (i.e. $\L'$ is in the fundamental Weyl chamber), 
and let $M'$ be a weight in the irreducible  weight representation $\cH_{\L'}$.
Then 
\begin{align}
 \langle\L'+\rho,\L'+\rho\rangle \geq  \langle M'+\rho,M'+\rho\rangle,
 \label{weight-inequality}
\end{align}
with equality if and only if $M'=\L'$.
Now consider the different possibilities $M' = M_i+L_i$ separately.
We choose $w\in \cW$ such that $\rho = w L_i$. 
Then $(\rho,\rho) = 2 = (L_i,L_i)$, and the Lemma implies 
\begin{align}
 \langle\L'+\rho,\L'+\rho\rangle  
  &\geq \langle w\cdot M'+\rho,w\cdot M'+\rho\rangle \nn\\
  &= \langle M_i+L_i+L_i,M_i+L_i+L_i\rangle 
  \label{estimate-S2}
\end{align}
for $M'=M_i+L_i$ a (dominant integral) weight in $\cH_{\L'}$. 
We conclude that the potential $V(\cA) = V_1(\cA) - V_2(\cA)$ is positive semi-definite.

\section{Vector zero modes}
\label{sec:app-zeromodes}

We want to identify the possible zero modes of the vector modes.
We use the same notation as in appendix \ref{sec:app-pos}.
Assume that $\cA$ is a zero mode at weight $M'$.
Up to an action with the Weyl group, we can assume that $M'$ is a dominant weight.
Then all the above inequalities must be equalities. This is not possible if $\cA$ is a superposition of
different $\L'$ representations, since then there is a strict
$>$ in  \eq{estimate}. Therefore $\cA=v_{\L'}$ is the highest weight vector in $\cH_{\L'}$.
Now expand $\cA_{\L'}$ as in \eq{Mprime-expand-tensor},
\begin{align}
 \cA_{\L'} &= \sum_i  \l_{L_i}\otimes  \cA^i_{M_i}  \ \ \in \ (8) \otimes \cH_\L  \  \nn\\
   L_i &\neq 0, \ \ M_i+L_i = M' = \L'
 \label{expandA-2}
\end{align}
Due to \eq{estimate-S2},  
$V_1[\cA_{\L'}] = V_2[\cA_{\L'}]$ can hold only if   
\begin{align}
  w_i\cdot (\L',L_i) &= (\L',\rho)  \nn\\   
   (\L',L_i) &= w_i^{-1} \cdot(\L',\rho)  
\end{align}
for all $i$ in this expansion, where $\L' = M_i + L_i$ is fixed.
This implies that $\w_i\L' = \L'$ for all $i$ in this expansion. 
Assume first that $\cW$ acts freely on $\L'$. Then  only one 
contribution with $L_i=\rho$ is allowed. Now the only highest weight vector in 
$(8)\otimes \cH_\L $ which has only one summand in \eq{expandA-2} is $v_{\L'}$
for $\L'=\L+\rho$, so that $M=\L$. This is the maximal weight representation in $(8)\otimes\cH_\L$. 
Taking into account the images under the Weyl group $\cW$, there are precisely 6 such zero modes.
We denote these as regular zero modes.
                          
Now assume that $\L'$ has a non-trivial stabilizer $w_2 \in \cW$,
\begin{align}             
 \L' = {w_2}^{-1}\cdot\L' 
\end{align}            
This is possible only for $\L' = m \L_1$ and $\w_2$ a reflection along $\a_2$, 
or $\L' = m \L_2$  and $\w_2$ a reflection along $\a_1$. 
Consider first the case $\L' = m \L_1$.  
Hence two and only two terms can occur in \eq{expandA-2}, with
$\L' = M_1+\rho =M_2+L_2$  for 
\begin{align} 
 L_2 &= w_2\rho = \rho - \a_2 = \a_1
 \end{align}
 and therefore
 \begin{align} 
 M_2 &=   M_1 + \rho - (\rho - \a_2) = M_1+\a_2 
\end{align}
The highest weight vector $\cA_{\L'}$ is therefore decomposed in the form
\begin{align}
 \cA_{\L'} &=  \l_\rho \otimes  \cA^i_{M_1} + \l_{w_2\rho } \otimes  \cA^i_{M_1+\a_2}
 \ \in  \   \cH_{\L'}  \subset (8)\otimes\cH_\L   \nn\\
\L' &= m \L_1 = M_1 + \rho = (M_1+\a_2) + \a_1 .
 \label{L'-decomp}
\end{align}
Now we expect that the only\footnote{Indeed for highest weights $\L' = \L + L_i$ 
with $L_i\neq \a_1,\a_2,\rho$, there are more than two ways
of writing $\L' = M_j+L_j$.
Unfortunately we do not have an airtight argument for the corresponding 
expansion \eq{expandA-2} of  the highest weight vectors $v_{\L'}$, so we leave it as a conjecture for now.} 
highest weight vectors in 
$(8)\otimes\cH_\L$ which have precisely two such summands are $v_{\L'}$
for $\L' = \L +\a_1$ and $\L' = \L +\a_2$. We focus on 
the first case here. Comparing with last term in \eq{L'-decomp} it follows that  
$\L  = M_1+\a_2$, and together with $\L' = m \L_1$ we obtain 
\begin{align}
 \L = \L' - \a_1 = m \L_1 - \a_1 = (m-2)\L_1+\L_2
\end{align}
since $\a_1 = 2\L_1-\L_2$.
Acting with the Weyl group on this $\cA_{\L'}$ gives three such zero modes\footnote{Recall 
that $\L'$ has a non-trivial stabilizer.}, which we shall call ``exceptional zero modes''.
Repeating the above argument for $\L' = m \L_2$ leads to 
analogous zero modes for $\L = \L_1+(m-2)\L_2$.

\end{document}